\documentclass[preprint,12pt]{elsarticle}

\usepackage{graphicx}
\usepackage{subfigure}
 \usepackage{hangcaption}
 \usepackage{amsmath,amssymb}

\begin{document}
 \def\half{ {1\over 2} }
 \def\tab{ {\hskip 0.15 true in} }
 \def\vtab{ {\vskip 0.1 true in} }
 \def\htab{ {\hskip 0.1 true in} }
  \def\ntab{ {\hskip -0.1 true in} }
 \def\vtabb{ {\vskip 0.0 true in} }
 \def\blah{ {\vskip 0.1 true in} }
 \def\Order{\mbox{$\cal{O}$}}
 \def\vec#1{ {\overline {#1}} }
 \def\vecc#1{ {\overline {#1}} }
 \def\congruent{=}

      \newcommand{\beq} { \begin{equation}  }
      \newcommand{\eeq}{ \end{equation} }
       \newcommand{\IR}{I \!\! R}
       \newcommand{\IS}{I \!\!\! S}
       \newcommand{\IZ}{Z}
       \newcommand{\sfrac}[2]{{\scriptstyle \frac{#1}{#2}}}
        \newcommand{\RR}{\bf R}
        \newcommand{\XX}{\bf X}
\newcommand{\bfomega}{\mbox{\boldmath $\omega$ \unboldmath} \hskip -0.075 true in}
\newcommand{\bfmu}{\mbox{\boldmath $\mu$ \unboldmath} \hskip -0.05 true in}
\newcommand{\xx}{{\bf x}}
\newcommand{\yy}{{\bf y}}
\newcommand{\bb}{{\bf b}}
\newcommand{\zz}{{\bf z}}
\newcommand{\qq}{{\bf q}}
\newcommand{\pp}{{\bf p}}
\newcommand{\bfxi}{\mbox{\boldmath $\xi$ \unboldmath} \hskip -0.075 true in}
\newcommand{\bfsigma}{\mbox{\boldmath $\sigma$ \unboldmath} \hskip -0.075 true in}
\newcommand{\bftau}{\mbox{\boldmath $\tau$ \unboldmath} \hskip -0.075 true in}
\newcommand{\bfepsilon}{\mbox{\boldmath $\epsilon$ \unboldmath} \hskip -0.075 true in}
\newcommand{\bfphi}{\mbox{\boldmath $\phi$ \unboldmath} \hskip -0.075 true in}
\newcommand{\bfchi}{\mbox{\boldmath $\chi$ \unboldmath} \hskip -0.075 true in}
\newcommand{\bfdelta}{\mbox{\boldmath $\delta$ \unboldmath} \hskip -0.075 true in}
\newcommand{\Real}{\ensuremath{\mathbb R}}
\newcommand{\SE}[1]{\ensuremath{{\mathrm{SE}(#1)}}}
\newcommand{\SO}[1]{\ensuremath{{\mathrm{SO}(#1)}}}
\newcommand{\bea}{\begin{eqnarray*}}
\newcommand{\eea}{\end{eqnarray*}}
\newcommand{\beaq}{\begin{eqnarray}}
\newcommand{\eeaq}{\end{eqnarray}}
\newcommand{\bz}{{\bf 0}}
\newcommand{\askip}{\htab\htab {\rm and} \htab\htab}
\newcommand{\asskip}{\htab {\rm and} \htab}
\newcommand{\wskip}{\htab {\rm where} \htab}
\newcommand{\wwskip}{\htab\htab {\rm where} \htab\htab}
\newcommand{\cX}{\tilde{X}}
\newcommand{\fracone}[1]{\frac{1}{#1}}
\newcommand{\se}[1]{\ensuremath{\mathfrak{se}(#1)}}
\newcommand{\so}[1]{\ensuremath{\mathfrak{so}(#1)}}
 \renewcommand{\vec}[1]{\boldsymbol{#1}}
 \newcommand{ \mps }[3]{ #1 \st #2 \rightarrow #3}
  \newcommand{\tldE}{{E}}
\newcommand{\tldS}{{S}}
\newcommand{\tlds}{\vec{s}}
\newcommand{\tldsh}{\widehat{\vec{s}}}
\newcommand{\bolda}{{\bf c}}
\newcommand{\bai}{c}
\newcommand{\trace}{ \mathit{\mathrm{trace}\,}}
  \def \st { \; { \mathbf :} \; }
\newcommand{\greekbf}[1]{\mbox{\boldmath$#1$}}

\renewcommand{\vec}[1]{{\bf #1}}
\newcommand{\mean}[1]{\left< #1 \right>}

\begin{frontmatter}



\title{A Modified Cross Correlation Algorithm for
Reference-free Image Alignment of Non-Circular Projections in Single-Particle Electron
Microscopy}


\author[wpark]{Wooram Park}
\ead{wooram.park@utdallas.edu}
\author[gchirik]{Gregory S. Chirikjian\corref{corres}}
\ead{gregc@jhu.edu}
 \cortext[corres]{Corresponding author,
Telephone : 1-410-516-7127, Fax : 1-410-516-4316}

\address[wpark]{Department of Mechanical Engineering, University of
 Texas at Dallas\\ Richardson, TX, 75080, USA  }

\address[gchirik]{Department of Mechanical Engineering, Johns
Hopkins University\\ Baltimore, MD, 21218, USA  }

\begin{abstract}
In this paper we propose a modified cross correlation method to
align images from the same class in single-particle electron
microscopy of highly non-spherical structures. In this new method,
first we coarsely align projection images, and then re-align the
resulting images using the cross correlation (CC) method. The
coarse alignment is obtained by matching the centers of mass and
the principal axes of the images. The distribution of misalignment
in this coarse alignment can be quantified based on the
statistical properties of the additive background noise. As a
consequence, the search space for re-alignment in the cross
correlation method can be reduced to achieve better alignment. In
order to overcome problems associated with false peaks in the
cross correlations function, we use artificially blurred images
for the early stage of the iterative cross correlation method and
segment the intermediate class average from every iteration step.
These two additional manipulations combined with the reduced
search space size in the cross correlation method yield better
alignments for low signal-to-noise ratio images than both
classical cross correlation and maximum likelihood (ML) methods.

\end{abstract}


\end{frontmatter}



\section{Introduction}
In single-particle electron microscopy, the main goal is to
reconstruct three-dimensional biomolecular complexes from noisy
planar projections obtained from transmission electron microscopes.
This structural information leads to better understanding of the
function and mechanisms of bio-macromolecular complexes. Since intensive
computation is required for this three-dimensional reconstruction,
faster and more accurate algorithms for reconstruction and
preprocess of two-dimensional images have been pursued extensively.
Several widely used computational packages have been developed for this purpose (e.g.
EMAN \cite{ludtke1999eman}, SPIDER \cite{shaikh2008spider}, IMAGIC
\cite{van1996new} and XMIPP \cite{sorzano2004xmipp})

In electron microscopy with biomolecular complexes, the electron
dose is limited to avoid structural damage on the specimen by
high-energy electrons. This leads to an extremely low
signal-to-noise ratio (SNR) in electron
micrographs~\cite{frank2006three}. One conventional approach to
deal with the low SNR images is to consider a {\it class} of
images corresponding to the same (or quite similar) projection
direction. Each image in a class can be thought of as the sum of
the same clear projection of the three-dimensional structure and a
random background noise field. A {\it class average} is the
representative image for each class. During the averaging process,
the additive background noise cancels and the resulting
average is a high SNR image and is believed to be close to the
clear projection. Prior to the class averaging, an alignment is
required to estimate the pose (position and orientation) of the
underlying projection in each image. Needless to say, more
accurate and faster algorithms for alignment will result in better
reconstruction results. The focus in this paper is on a method that is particularly
well suited to non-spherical particles such as ion channels. The projections
of these non-spherical particles are typically non-circular, leading us to investigate how
to exploit this anisotropy to improve existing class-averaging algorithms. Before
discussing our approach, a brief review of existing methods is given below.

The cross correlation (CC) method is one of the
most popular computational tools for this
problem~\cite{penczek1992three}. The maximum cross correlation
occurs at the best alignment of two images. However, if the
SNR of images is low, false peaks in the cross correlation
function degrade the accuracy of the cross correlation method.
More recently, Penczek et al.~\cite{yang2008cryo} proposed a new
alignment method using nununiform FFT. They use a gridding method
to re-sample images with high accuracy, and then find a better
alignment for the images. The computational efficiency of various
alignment methods was also investigated
in~\cite{joyeux2002efficiency}.

An alternative to the CC approach is the maximum likelihood (ML)
method developed in~\cite{sigworth1998maximum}. This method
does not find the alignment for each image in a class directly.
Rather it finds the underlying projection using statistical models
for the background noise and the pose of the projection. The
likelihood is defined as a function of the projection image and
the parameters for the statistical models. The {\it refinement}
process finds the projection image and the parameters by
maximizing the likelihood function. This approach has been
extended to deal with the case where data images of a class are
heterogeneous~\cite{scheres2005maximum}.

Typically both the CC and ML methods are implemented as an
iterative process and require an initial guess for the underlying
projection image~\cite{sigworth1998maximum}. Due to this
requirement, users should intervene in the computational process.
If the preliminary structural information (e.g. symmetry, low
resolution features, etc.) of the biological complex of interest
is given, it is relatively easy to choose the initial image for
the iteration. However, this is not the case if the biological
complex is being studied for the first time. Moreover, even if
some preliminary information about the structure is given, it is
still a hard problem to choose the best starting image. And all
other things being equal, a method that does not require human
intervention is inherently better than one that does.

In the conventional CC method, all possible alignments are
searched. In other words, the CC of two images is computed as a
function of relative translations and rotations, and then the
optimal alignment maximizing the CC is chosen. To search the
optimal translation, the discrete Fourier transform (DFT) is a
useful and fast tool~\cite{penczek1992three,
joyeux2002efficiency}. However, to search the optimal rotation, an
image is rotated by every possible rotation angle, and then the CC
with the other image is computed. For an asymmetric projection
image, a search of angles from 0 to $2\pi$ is required. In
addition, limited resolution due to discretization of angles is
inevitable. Since the rotation involves computationally expensive
interpolation, a fine discretization increases computation time,
even though it may give better accuracy.

Penczek et al.\cite{penczek1992three} proposed a reference-free
alignment algorithm. It consists of two steps: 1) ``random
approximation'' of the global average, and 2) refinement with the
result from the first step. In the first step, images are
sequentially aligned and averaged in randomized order. In the
second step, the alignment for each image from the first step is
improved so that each image is best aligned to the average of the
rest of images. Marco et al.~\cite{marco1996variant} modified the
first step to avoid the effect of the order of input images. They
proposed a pre-alignment method based on a pyramidal structure,
instead of the sequential alignment. All the images are paired,
aligned and averaged. Then
  the same process is repeated to the resulting images until one
image remains.

Often ML outperforms CC. However, in this paper we explore a modification
to CC for non-spherical particles that significantly improves its performance.
Namely, we pre-align classified
images and then apply the CC method to re-align the class
images\footnote{We assume that an initial classification is made
by an existing algorithm such as EMAN~\cite{ludtke1999eman}.
Recent {\it classification-free} methods presented in
\cite{coifman2010reference} are another possible alternative to
existing algorithms.}. During the pre-alignment, the images are
coarsely aligned by matching the centers of mass and the principal
axes of images. The second step (re-alignment) uses the resulting
average, the alignment and the distribution of misalignment from
the first step (pre-alignment). The most important benefit of this
pre-alignment is that we can estimate the pose distribution of the
misalignment. This distribution enables us to reduce the search
space for the CC method to those poses that are most probable.
Since the search space is reduced, the
sampling interval is also reduced for a specified number of
samples. Using synthetic data images, we show that our new method
produces better results than both the conventional CC and ML methods.

The remainder of this paper is organized as follows. In Section
\ref{sc:review}, we review two existing methods (the cross
correlation and maximum likelihood methods) for class averaging in
single particle electron microscopy. In Section
\ref{sc:new_method}, we propose a new method to better pre-align very noisy
images which has a pose distribution for misalignment that has a closed analytical form.
In Section \ref{sc:result}, the results obtained by the
new and existing methods are presented and the resulting images
are assessed using several measurement methods. Finally, the
conclusion is presented in Section \ref{sc:conclusion}.

\section{Review of the cross correlation and maximum likelihood methods} \label{sc:review}
A class average can be defined as
$$ \gamma(\vec{x}) = \frac{1}{N}
\sum_{i=1}^{N} \rho_i(g_i^{-1} \cdot \vec{x}),
$$
where $\rho_i(\vec{x})$ is the $i^{th}$ image in a class and $g_i
= g({\bf q}_i)$ represents the planar rigid-body motion
responsible for alignment of the image with roto-translation
parameters\footnote{This notation ${\bf q}$ for roto-translation
parameters is corresponding to $\phi$ in
\cite{sigworth1998maximum}. We use this since this variable
denotes a vector.} ${\bf q}_i = (q_{\theta_i}, q_{x_i}, q_{y_i})$.
In this context, each rigid-body transformation such as $g_i$ can
be thought of as a particular evaluation of the matrix-valued
function $g({\bf q})$ defined as
\begin{equation}\label{deflwlekwd}
g({\bf q}) = \left(\begin{array}{ccc}
\cos q_\theta & -\sin q_\theta & q_x \\
\sin q_\theta & \cos q_\theta & q_y \\
0 & 0 & 1 \end{array}\right). \end{equation} Moreover, each $g_i$
performs the ``action,'' $\cdot$\,, of moving a point in the
plane, $\vec{x} \in \mathbb{R}^2$. The optimal alignment can be
obtained by maximizing the following
quantity~\cite{penczek1992three}:
\begin{equation}\label{eq:costfunction}
C(g) =  \left| \sum^{N}_{i=1} \rho_i(g_i^{-1} \cdot \vec{x})
\right|^2.
\end{equation}
It was shown in previous publications that this problem can be
solved using iterative optimization. After the $n^{th}$ iteration,
the next iteration result is given as~\cite{sigworth1998maximum}
\begin{equation}\label{eq:argmax}
g_i^{(n+1)} = \arg\max_{\hspace{-5mm}g} \left( \rho_i(g^{-1} \cdot
\vec{x})\odot \left[ \gamma^{(n)}(\vec{x}) - \frac{1}{N}   \rho_i(
g^{-1} \cdot \vec{x} ) \right] \right) , \hspace{5mm} i=
1,2,\cdots N,
\end{equation}
where $\odot$ denotes the inner product between two image arrays, such
that
$$
A \odot B = \sum_{k,l} a_{kl} b_{kl}.
$$
Using the improved alignment $g^{(n+1)}$, the averaged image is
refined as
$$
\gamma(\vec{x})^{(n+1)} = \frac{1}{N} \sum_{i=1}^{N} \rho_i\left(
\left[g_i^{(n+1)}\right]^{-1} \cdot \vec{x}\right).
$$

To find the maximizer in (\ref{eq:argmax}), the cross correlations
for possible alignments (translations and rotations) are computed
and the maximizer is chosen. One image is actually rotated by
candidate rotation angles and the cross correlation of the two
images are computed as a function of translation. This can be
easily implemented using the discrete Fourier transform. For
various rotation angles, we stack the cross correlation and the
three-dimensional search for the maximum CC gives the optimal
alignments. This alignment method is referred to as {\it direct
alignment using 2D FFT} in \cite{joyeux2002efficiency}.

The image rotation of discrete images requires interpolation.
Since every class image should be rotated several times by
possible rotation angles, the computation time for the whole class
images is considerable. There is a trade-off between the
computation time and the accuracy of the result. In addition, the
CC method fails with low SNR images, because of the existence of
false peaks in the cross correlation.

The maximum likelihood (ML) method for image refinement in
single-particle electron microscopy shows better performance than
the CC method, especially for low SNR
images~\cite{sigworth1998maximum}. The ML method defines the
likelihood function based on a statistical model for the additive
background noise and the pose (position and orientation) of the
underlying clear projection relative to the bounding box.
In~\cite{sigworth1998maximum},  Gaussian distributions are used to
describe both the background noise and the positional distribution
of the projection along $x$ and $y$ axes. The rotational angles of
the projection are assumed to be uniformly distributed.

The ML method for image refinement maximizes the following
function~\cite{sigworth1998maximum}:
\begin{equation}\label{eq:L_func}
L(\Theta) = \sum_{i=1}^{N}\log \int P(\rho_i|{\bf q},\Theta)f({\bf
q}|\Theta) d{\bf q}
\end{equation}
where
\begin{equation}\label{eq:p_exp}
P(\rho_i|{\bf q},\Theta) = \left( \frac{1}{\sqrt{2\pi}\sigma}
\right) \exp\left(-\frac{|\rho_i({\bf q})-A|^2}{2\sigma^2}
\right),
\end{equation}
and
\begin{equation}\label{eq:f_exp}
f({\bf q}|\Theta)d{\bf q} = \frac{1}{2\pi \xi_{\sigma}^2}
\exp\left[
-\frac{(q_x-\xi_x)^2+(q_y-\xi_y)^2}{2\xi^2_{\sigma}}\right]
\frac{dq_{\theta}}{2\pi}dq_x dq_y.
\end{equation}
As before, ${\bf q}=(q_{\theta},q_x,q_y)$ represents the
parameters defining a rigid-body transformation (rotation and
translations), and $\Theta$ denotes the underlying projection
($A$), the standard deviation ($\sigma$) of the background noise,
and the mean $(\xi_x,\xi_y)$ and the standard deviation
$(\xi_{\sigma})$ of the positions of the centers of mass of the
projections.

The probability density in (\ref{eq:p_exp}) is defined based the
assumption that the background noise is Gaussian with the
variance, $\sigma^2$. As shown in (\ref{eq:f_exp}), the
distributions of the $x$ and $y$ positions of the center of mass
of the projection in a class are modeled as a Gaussian with the
mean position $(\xi_x,\xi_y)$ and the variance, $\xi_{\sigma}^2$.
While the Gaussian assumption for the background noise is widely
accepted, the Gaussian function in (\ref{eq:f_exp}) should be
rationalized carefully. Moreover, for projections of non-spherical
particles resulting in non-circular images, one would expect that
a $f({\bf q}|\Theta)$ which depends in some way on $q_{\theta}$
would be more informative than one that does not. An exact form
for this kind of dependence is given later in the paper.

In single-particle electron microscopy in general, many projections in a
large micrograph are selected by a particle selection program with
a bounding box (e.g. {\it boxer} in EMAN~\cite{ludtke1999eman}).
The small images containing one projection with the additive noise
are then grouped into classes. The two-dimensional positions and
one-dimensional rotation of the underlying projection relative to
the bounding box can be assumed to be Gaussian as
in~\cite{sigworth1998maximum}. However, an analysis of the
statistical behavior of the particle selection should be performed
first. Even if the assumption about the distribution is
acceptable, the ML method intrinsically requires the integral over
the two-dimensional translations and the one-dimensional rotation
in (\ref{eq:L_func}). As in the CC method, the discretization of
the rotation angle is an issue. Using more angular samples
requires more computation time as a price for a potentially more
accurate solution.

To implement the ML method, an iterative update for the underlying
projection is used. The $(n+1)^{th}$ iteration for the underlying
projection after the $n^{th}$ iteration is given
as~\cite{sigworth1998maximum}
\begin{equation}\label{eq:ML_A}
A^{(n+1)}=\frac{1}{N}\sum_i \frac{\int \rho_i({\bf q}) r_i({\bf
q};\Theta^{(n)})d{\bf q}}{\int r_i({\bf q};\Theta^{(n)})d{\bf q}},
\end{equation}
where
$$
r_i({\bf q};\Theta) =P(\rho_i|{\bf q},\Theta)f({\bf q}|\Theta).
$$
This iterative process requires an initial starting image, $A^{0}$.
Since the function $r_i(\cdot)$ can be rewritten as
$$
r_i({\bf q};\Theta) = k_i \exp\left( \frac{\rho_i({\bf q})\cdot
A}{\sigma^2} \right)f({\bf q}|\Theta),
$$
the cross correlation should be computed in the ML method.

As we briefly reviewed here, the iteration process in the CC and
ML methods requires a reference as a starting image. Even though a
reference-free alignment method is
available~\cite{penczek1992three}, it is essentially a two-step
method; the first step generates a reference image out of data
images and then the second step refines the reference iteratively.
In addition to the issue about reference images, the cross
correlation is computed for various alignments to find the maximum
CC or implement the integral over alignments. A finer
discretization for the rotation angles may yield better accuracy,
but this comes at the cost of increased computation time.

\section{Methods}\label{sc:new_method}
The new method proposed in this paper consists of two parts:
pre-alignment of class images and application of the CC method to
the pre-aligned images with blurring and segmentation.

\subsection{Matching centers of mass and principal axes of images}\label{sc:CMPA}
Matching the centers of mass and the principal axes (CMPA) of two
images gives the alignment of a class of
images~\cite{park:ijrr_stoch}. The accuracy of the alignment by
this method is sensitive both to the background noise and the degree of circularity of the underlying pristine projection.
However, the advantage of this alignment method is that we can quantify the
distribution of the misalignments. This provides a
better starting point than assuming a uniform orientation
distribution.

As derived in~\cite{park:ijrr_stoch}, the probability density
function for the misalignments after the CMPA matching is given as\footnote{In that paper,
a method for resolving the 180-degree ambiguity in principal-axis alignment was also provided
to make the resulting orientational distribution unimodal in cases of relatively high SNR (e.g., 0.2 and higher). But this
symmetry-breaking fails for case of low SNR (e.g., 0.05 and lower) and the statistical characterization
of this in a way that can be used in CC is nontrivial, and so the version of $p(\cdot\,;\,\cdot)$ used here
is bimodal.}
\begin{equation}\label{eq:pdf_misalignment}
p(q_x,q_y,q_{\theta};\xi_{\sigma},\xi_{\theta}) = \frac{1}{8\pi^2
\xi_{\sigma}^2} e^{ -(q_x^2+q_y^2)/(2\xi^2_{\sigma}) }
\left(\sum_{k=-\infty}^{\infty} e^{-\frac{k^2 \xi_{\theta}^2}{2}}
e^{i k q_{\theta}}+\sum_{k=-\infty}^{\infty} e^{-\frac{k^2
\xi_{\theta}^2}{2}} e^{i k (q_{\theta}-\pi)} \right).
\end{equation}
While the misalignments of translation forms a unimodal Gaussian
distribution, the misalignments of rotation forms a bimodal
distribution. This is because an image has two equivalent
principal axes whose directions are opposite to each other. Though
this ambiguity makes it difficult to determine the rotational
alignment, it is easy to have the resulting distribution for the
rotational misalignment. It is essentially the sum of two Gaussian
functions wrapped around the circle with the same standard deviation
$\xi_{\theta}$ and two different means, 0 and $\pi.$ For $N\times
N$ images, the parameters in (\ref{eq:pdf_misalignment}) are
computed directly from the background noise properties as~\cite{park:ijrr_stoch}
\begin{equation}\label{eq:sigma_define1}
\xi_{\sigma} = \sqrt{K \sum_{l=1}^N x_l^2}
\end{equation}
\begin{equation}\label{eq:sigma_define2}
\xi_{\theta} = \sqrt{   \frac{K}{ (\lambda_1-\lambda_2)^2 }
    \left( K \sum_{l=1}^N x_l^2\sum_{k=1}^N y_k^2  + \sum_{l=1}^N x_l^2 y_l^2
    \right) }
\end{equation}
where $x_l=y_l=l-(N+1)/2$, $K = (1+4\nu)\sigma^2/M^2$, $\sigma^2$
is the variance of the background noise, and $\nu $ is the
correlation coefficient between the noise in adjacent pixels. $M$
is defined as $M =   \frac{1}{N}\sum_i\sum_j \rho_i(\vec{x}_j) $,
which is the mean of the sum of the pixel values of images. The
sums in (\ref{eq:sigma_define1}) and (\ref{eq:sigma_define2}) can
be simplified as closed-form expressions as
$$
\xi_{\sigma} =  \sqrt{\frac{K}{12} N(N^2-1) }
$$
$$
\xi_{\theta}  = \sqrt{   \frac{K}{ (\lambda_1-\lambda_2)^2 }
    \left( \frac{K}{144} N^2(N^2-1)^2 + \frac{N}{240}(3N^2-7)(N^2-1) \right)
    }.
$$
The inertia matrix of an image aligned by matching CMPA is
computed as
$$
J_i = \frac{1}{M}\sum_j \vec{x}_j\vec{x}_j^T \rho_i'(\vec{x}_j ) =
\left(\begin{array}{cc}
  L_x(i) & 0 \\
  0 & L_y(i) \\
\end{array}\right).
$$
Note that the image $\rho_i'(\vec{x}_i )$ is a version of $\rho_i(\vec{x}_i )$ which is
aligned so as to have a diagonal inertia matrix. The term $(\lambda_1-\lambda_2)$ is
defined as
$$
(\lambda_1-\lambda_2)=\frac{1}{N}\sum_i (L_x(i)-L_y(i)).
$$
Note that as $|\lambda_1-\lambda_2| \, \rightarrow \, 0$, as would
be the case for a circular image, $ \xi_{\theta} \, \rightarrow \,
\infty$, and the folded normal reduces to the uniform distribution
on the circle. This may not be obvious from the form given in
(\ref{eq:pdf_misalignment}), but by writing this same
orientational distribution in the form of a Fourier series as is
done in Eq.~2.46 in \cite{chirikjian2009stochastic}, the
convergence to uniformity as $\xi_{\theta}$ becomes infinite
becomes obvious. In this case $p({\bf
q};\xi_{\sigma},\xi_{\theta})$ reduces to a form akin to $f({\bf
q}|\Theta)$ in (\ref{eq:f_exp}). Hence, the method used here is
general, though the value that it adds to the existing literature
are realized when the projections are anisotropic and hence the
smaller $\xi_{\theta}$ is, the more useful our approach becomes.

General approaches to compute the alignment error in data images
are developed in ~\cite{jensen2001alignment} and
\cite{baldwin2005estimating}. In our case, we aimed to
characterize the alignment error which the specific alignment
method (the CMPA matching) produces.

This matching algorithm has one more benefit compared to the
reference-free alignment in \cite{penczek1992three} and
\cite{marco1996variant}. In the CMPA matching method, each image
can be aligned independently, while two images should be
considered to align in \cite{penczek1992three} and
\cite{marco1996variant}. Essentially we align images to a
reference frame in the CMPA matching. In other words, the center
of mass and the principal axis of a image is matched to a
space-fixed reference frame rather than pairwise between images.
Therefore the alignment result is not dependent of the order in
which we consider the input images. In contrast, the first step of
the reference-free method in \cite{penczek1992three} is dependent
on the input order. Even though Marco et
al.~\cite{marco1996variant} developed an alternative method which
is less sensitive to the input order, it is not completely
independent of the order.

Obviously, with high SNR images, matching the CMPA of images will
generate accurate alignment. In this case the misalignment can be
removed from a blurry class average using a deconvolution
technique \cite{park2010deblurring}. For low SNR images, we will
apply a new method which we propose in the next subsection.

\subsection{Modified CC method}

\subsubsection{Search space for alignment}
As seen in Section \ref{sc:CMPA}, the statistics of misalignments
after the CMPA matching can be modeled using Gaussian functions
with the parameters defined in (\ref{eq:sigma_define1}) and
(\ref{eq:sigma_define2}), though we cannot compute the true
alignment for each image.

This reduces the search space. Without the CMPA matching approach,
the search space for rotation would be $[0, \,\, 2\pi)$ and the
sampling interval should be equally spaced because there is no
information about the tendency of orientation. However, if we use
the pre-aligned images, we know that the true rotation angles
exist around the values 0 and $\pi$ with the computed standard
deviation. Thus, we can focus on a smaller search space.
Furthermore the sampling interval should be designed according to
the distribution. This sampling can be performed using inverse
transform sampling. A sample value $X$ is obtained as
$$
X = F_c^{-1}(Y)
$$
where $F_c$ is the cumulative density function and $Y$ is drawn
from a uniform distribution on $(0, \,\,\,1)$.

\subsubsection{Image blurring and segmentation}
As is widely known, the CC method exhibits false peaks for low SNR
images. To avoid false maxima, we artificially blur the images
during the early iterations of the CC method. Practically we
convolve data images with a two-dimensional Gaussian to generate
the blurred version of the images. The method to choose the
optimal blurring parameter will be proposed in Section
\ref{sc:flow} (See Phase 2 in Figure \ref{fig:diagram}).

Since class images contain one projection of a single particle, we
can expect that there are two regions in the image: projection
image region and noise region. When we apply the CC method, the
background noise in the intermediate average
($\gamma^{(n)}(\vec{x})$ in (\ref{eq:argmax})) degrades the
performance of the CC method. This background noise can be
eliminated by a image segmentation technique, because it is easier
to distinguish the projection region and the noise region in the
intermediate average. We apply the edge detection algorithm
developed in~\cite{canny1986computational} to solve this
segmentation problem.

\subsubsection{Successive
transformations}\label{sc:successive_transformations}

The new method proposed here consists of the pre-alignment by CMPA
matching and the re-alignment by the iterative CC method with the
reduced search space. During the process, each image will be
repeatedly transformed (rotation and translation) to find the best
alignment. If we apply multiple transformations (rotations and
translations) on a two-dimensional discrete image successively,
the resulting image will have many artifacts since such
transformations of digital images involve interpolation. To
overcome this, instead of storing the transformed images for the
next iteration, we record the transformation information for each
image maintaining the original images. Two consecutive rigid body
 transformations on the plane result  in one transformation. The combined
transformation can be computed using the rigid body motion group
which is one popular mathematical tool in
robotics~\cite{chirikjian2001engineering}.

Two $3\times 3$ matrices representing rotation and translation on
the plane can be written using (\ref{deflwlekwd}) respectively as
$$
g_r(\theta)= \left(%
\begin{array}{cc}
  R(\theta) & \vec{0} \\
  \vec{0}^T & 1 \\
\end{array}%
\right) = g(\theta,0,0) \hspace{10mm}
g_t(\vec{p})= \left(%
\begin{array}{cc}
  I & \vec{p} \\
  \vec{0}^T & 1 \\
\end{array}%
\right) = g(0,p_1,p_2)
$$
where
$$
R(\theta) =  \left(%
\begin{array}{cc}
  \cos\theta & -\sin\theta \\
  \sin\theta & \cos\theta  \\
\end{array}\right),%
$$
${\bf 0}$ is the 2D zero vector, and ${\bf 0^T}$ is its transpose.

$g_r(\theta)$ and $g_t(\vec{p})$ represent pure rotation and
the pure translation in the plane, respectively. If we translate
and then rotate an image respectively by $\vec{p}$ and by $\theta$
relative to the frame of reference fixed at the origin, then the resulting
transformation is written as
$$
g'(\theta,\vec{p}) = g_r(\theta) g_t(\vec{p})=\left(%
\begin{array}{cc}
  R(\theta) & R(\theta)\vec{p} \\
  \vec{0}^T & 1 \\
\end{array}%
\right).
$$
Here the $'$ is used to distinguish this transformation from
$g(\theta,\vec{p}) = g_t(\vec{p}) g_r(\theta) = g(\theta,p_1,p_2)$.

Two successive transformations,
$g'(\theta_1,\vec{p}_1)$ followed by $g'(\theta_2,\vec{p}_2)$, can
be written as
$$
  g'(\theta_2,\vec{p}_2) g'(\theta_1,\vec{p}_1) = \left(%
\begin{array}{cc}
  R(\theta_2)R(\theta_1) & R(\theta_2)R(\theta_1)\vec{p}_1+R(\theta_2)\vec{p}_2 \\
  \vec{0}^T & 1 \\
\end{array}%
\right)
$$
\begin{equation}\label{eq:synthesis}
 =\left(%
\begin{array}{cc}
  R(\theta_2)R(\theta_1) & \vec{0} \\
  \vec{0}^T & 1 \\
\end{array}%
\right)\left(%
\begin{array}{cc}
  I & \vec{p}_1+R(-\theta_1)\vec{p}_2 \\
  \vec{0}^T & 1 \\
\end{array}%
\right)
\end{equation}
Therefore, the successive transformations can be viewed as the
translation by $\vec{p}_1+R(-\theta_1)\vec{p}_2$ followed by the
rotation by $(\theta_1+\theta_2)$. Note that all the
transformations here are performed using the fixed frame of
reference attached to the center of the bounding box.

Combined with the reduced search space, this tool enables a search
with finer alignment angles. In the conventional CC method and the
ML method, only the predefined discrete angles are considered.
Especially for the CC method, each class image is eventually
assigned to one of the predefined discrete angles. Since the
angles are equally-spaced samplings from $[0, \,\,2\pi)$, the
resolution of the rotational alignment is limited by $2\pi/N$,
where $N$ is the number of samplings. However, in our method, the
pre-alignment by the CMPA matching gives the arbitrary alignment
angles and the candidate alignment angles for re-alignment are
sampled within a smaller and more targeted search space guided by
knowledge of the mean and variance of the CMPA. During the
iteration, the re-alignment information for each image is obtained
and then the new combined transformation is computed using the
previous alignment information and the new alignment information.
We do not store the transformed images, rather store the alignment
information keeping the original class images. Using this
manipulation, we can avoid the image artifacts that may be caused
by multiple transformations.

\subsection{Flow of the new method}\label{sc:flow}
\begin{figure}
  \centering
   \includegraphics[width=5.5in]{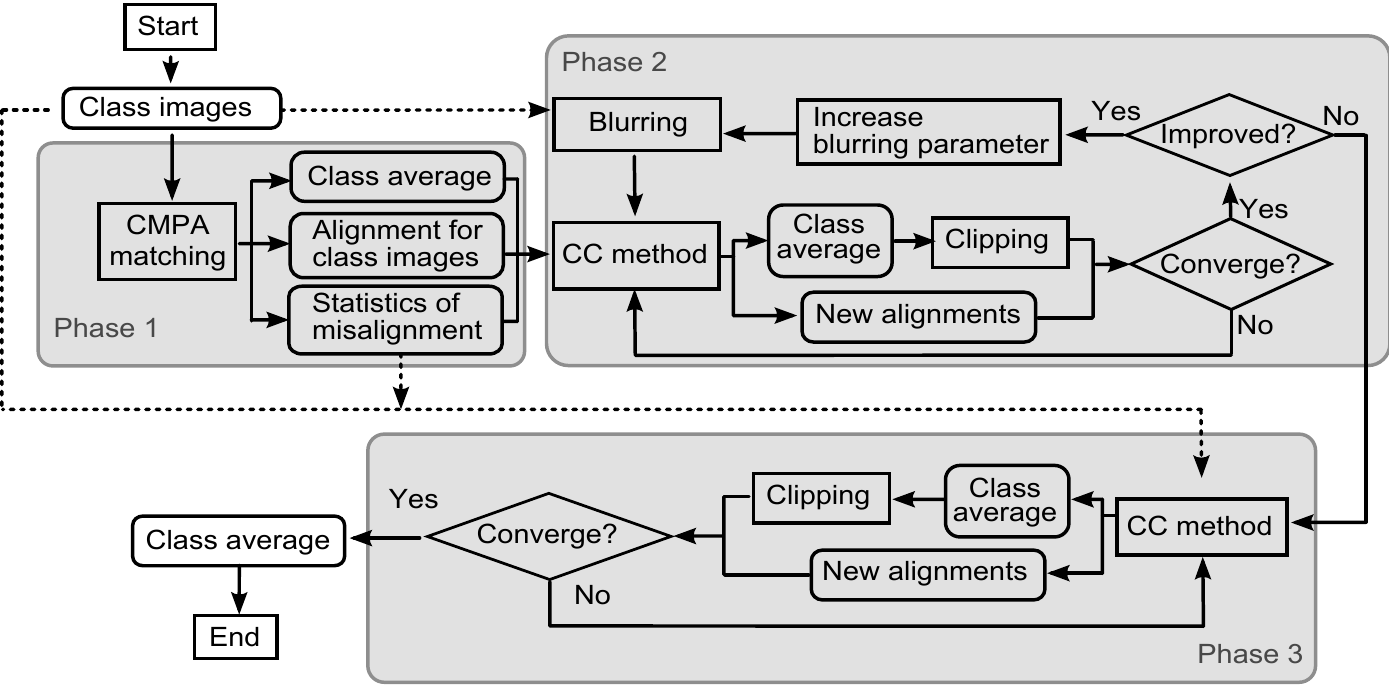}
  \caption{Diagram for the new alignment method. }\label{fig:diagram}
\end{figure}

The flow chart for the new alignment method is shown in Figure
\ref{fig:diagram}. The rectangles and the rounded rectangles
denote operations and data, respectively. The continuous lines
with arrows denote the main flow of the new method. The dotted
lines with arrows describes that the original images are used in
the subroutines.

In Phase 1, the images are coarsely aligned by matching the CMPA
of images. After this process, we have the alignment for every
image, an averaged image, and the statistical information about
misalignment involved in the coarse alignment.

In Phase 2, we first blur the images from Phase 1 using a Gaussian
kernel. We start with the standard deviation 0.25 pixel for the
Gaussian kernel. Then we apply the CC method to re-align the
blurred image. The iterative process in Phase 2 takes the averaged
image from Phase 1 as a reference image. Also the reduced search
space for alignments based on the distribution of misalignment is
applied. This iteration is repeated until it converges with 3\%
threshold. In other words, this iteration will stop when the image
improvement measured by the normalized lease-square error (NLSE)
is less than 3\%. After this iteration denoted by the lower loop
in Phase 2 in Figure \ref{fig:diagram}, we compute the cost
function (\ref{eq:costfunction}) to measure the effectiveness of
the artificial blurring. We repeat the lower loop iteration in
Phase 2 with the increased blurring parameters until we find the
optimal blurring parameter. The parameter is increased by 0.25
pixel for each step. This simple search for the blurring parameter
is valid because of the fact that the alignments without blurring
and with a large blurring will both produce bad results and the
optimal blurring parameter will exist in between. The re-alignment
in Phase 2 cannot be accurate because the blurred images are used.
Even though the re-alignment is not satisfactory, this process
gives better alignment than Phase 1 and we can avoid the problems
associated with false peaks in cross correlations.

In Phase 3, we find more accurate alignment. This phase apply the
CC method to the original version of images. The reduced search
space and the resulting alignments (from Phase 2) for images play
an important role in this phase. Iterations are performed until
they converge.

In Phase 2 and 3, the projection region in the averaged image
after each rotation is obtained using image segmentation in order
to avoid the effects of the noise surrounding the region of
interest in the image on the next iteration. In addition, we do
not store the rotated and translated images for the next
iteration. Rather, we use the original images with their alignment
information for the next iteration as denoted by the dotted lines
with the arrows. This reduces the interpolation error which may
occurs during repeated rotation and translation of images. For
given successive transformations, we can use a combined
transformation from the method in
Section~\ref{sc:successive_transformations}.

\begin{figure}
  \centering
   \subfigure[]{\label{fig:original}
   \includegraphics[width=1.6in]{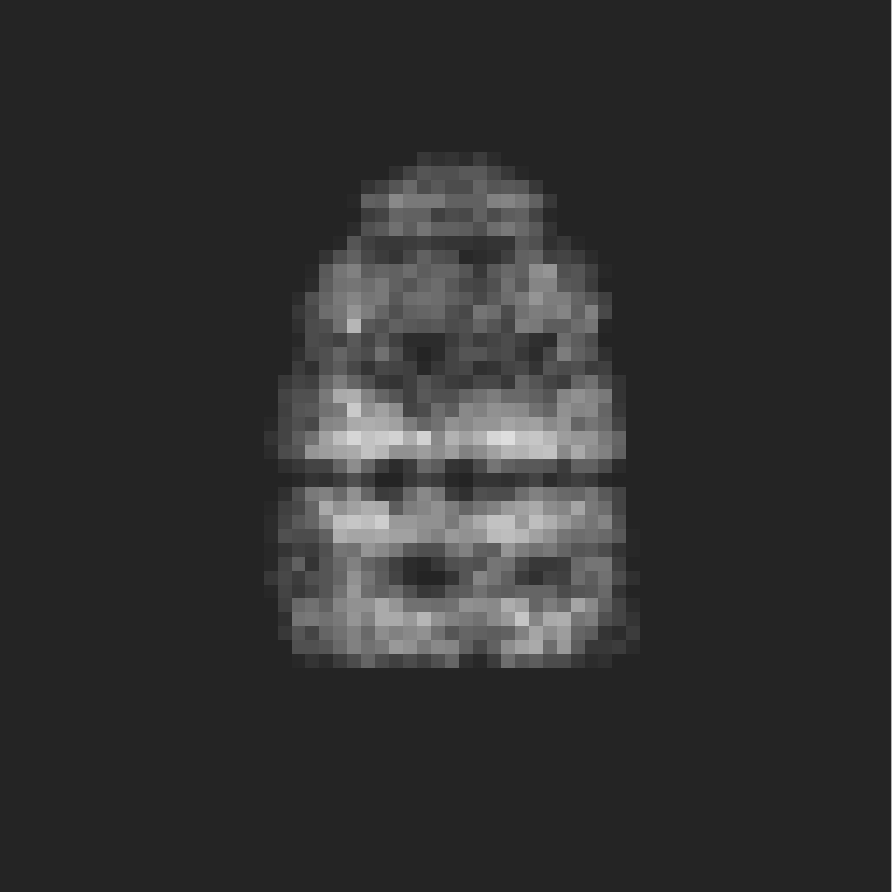}
}
 \subfigure[]{\label{fig:perfect_average}
  \includegraphics[width=1.6in]{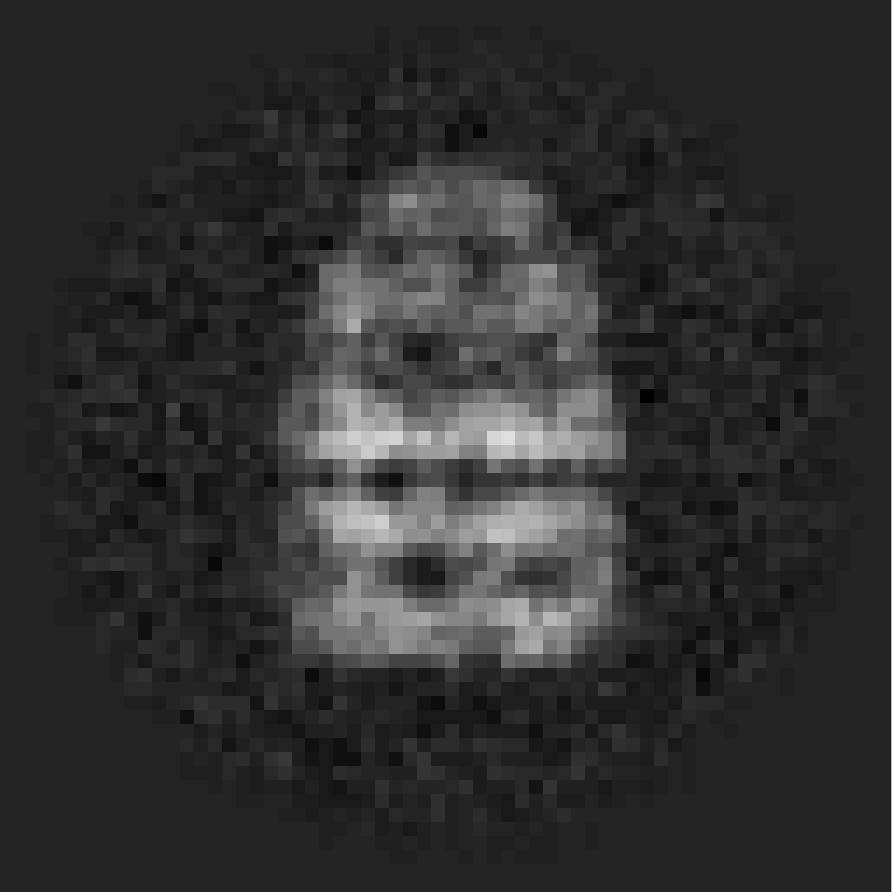}
  }
  \caption{(a) The original clear projection.
  (b) The average image with the unattainable perfect alignment for Case 1. }\label{fig:ori_ave}
\end{figure}

\begin{figure}
  \centering
 \subfigure[SNR=0.025, $\nu=0$]{\label{fig:noise_case1}
   \includegraphics[width=1.4in]{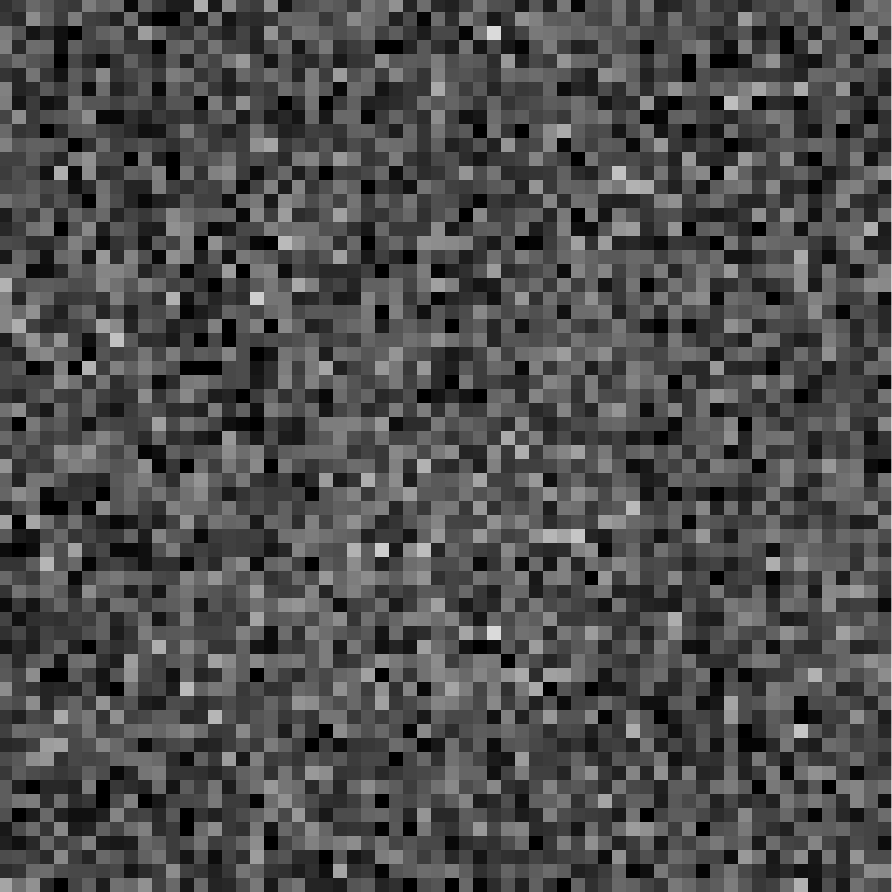}
}
 \subfigure[Blurred version of (a)]{\label{fig:blur_case1}
   \includegraphics[width=1.4in]{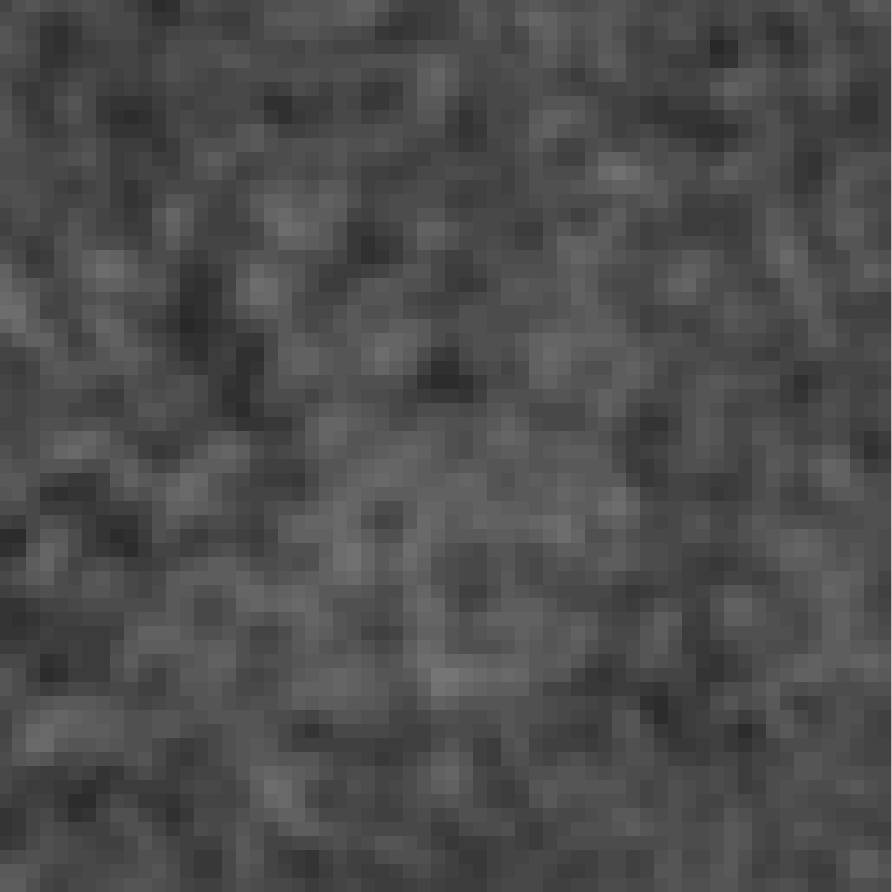}
}\\
 \subfigure[SNR=0.050, $\nu=0$]{\label{fig:noise_case2}
   \includegraphics[width=1.4in]{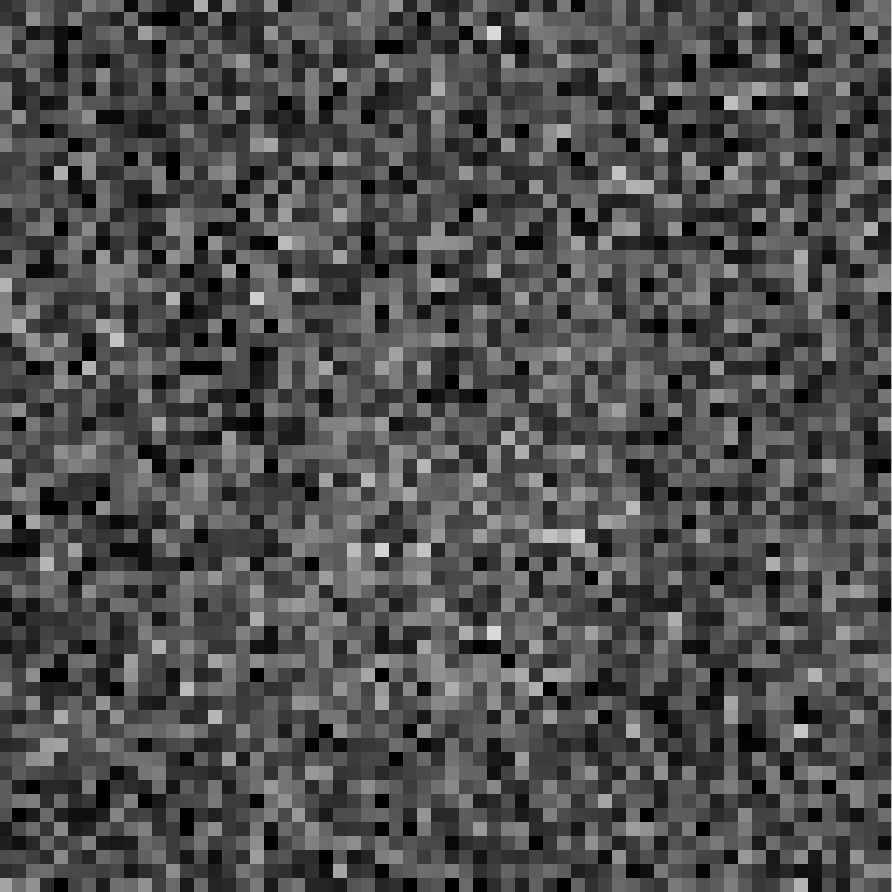}
}
 \subfigure[Blurred version of (c)]{\label{fig:blur_case2}
   \includegraphics[width=1.4in]{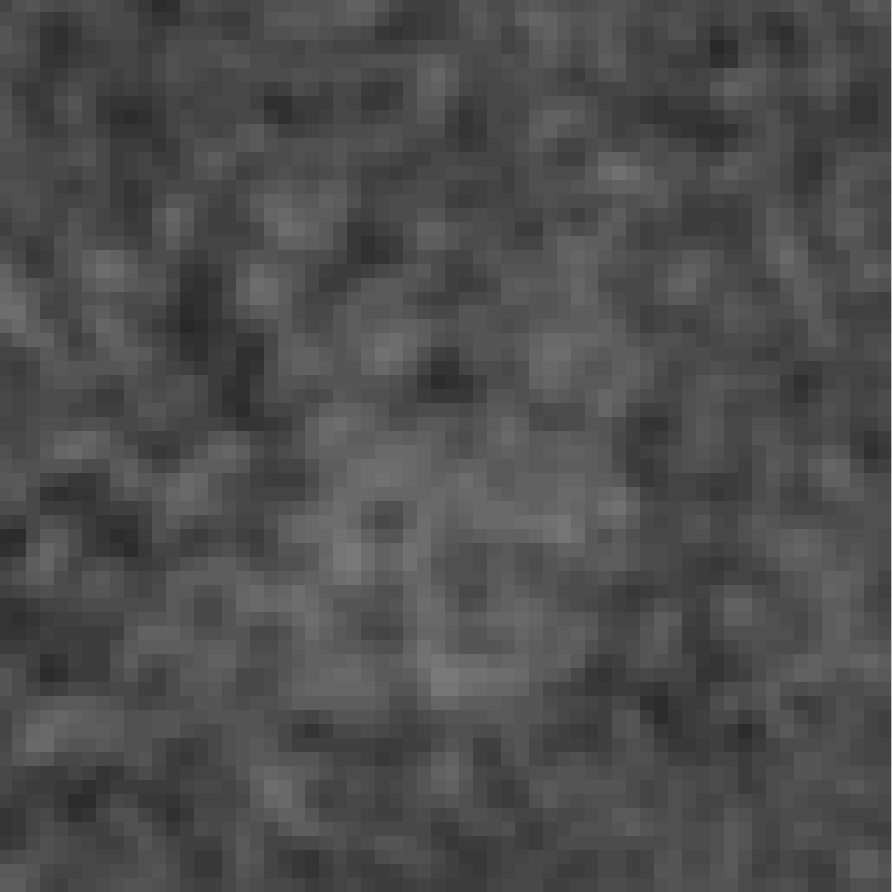}
}\\
 \subfigure[SNR=0.100, $\nu=0$]{\label{fig:noise_case3}
   \includegraphics[width=1.4in]{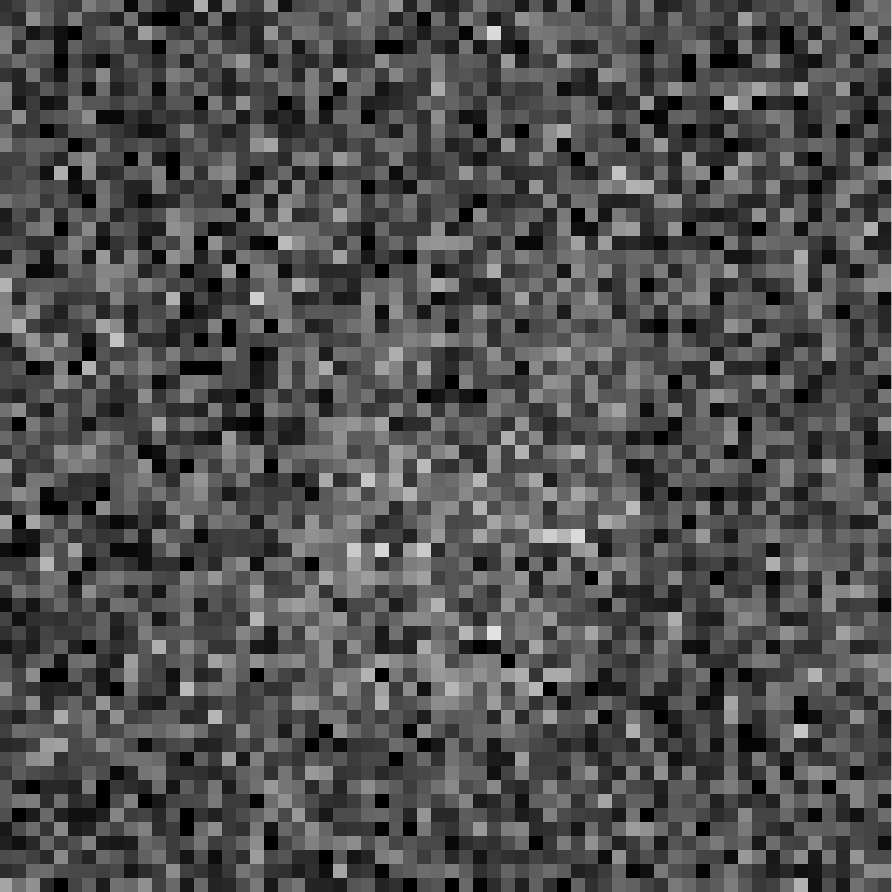}
}
 \subfigure[Blurred version of (e)]{\label{fig:blur_case3}
   \includegraphics[width=1.4in]{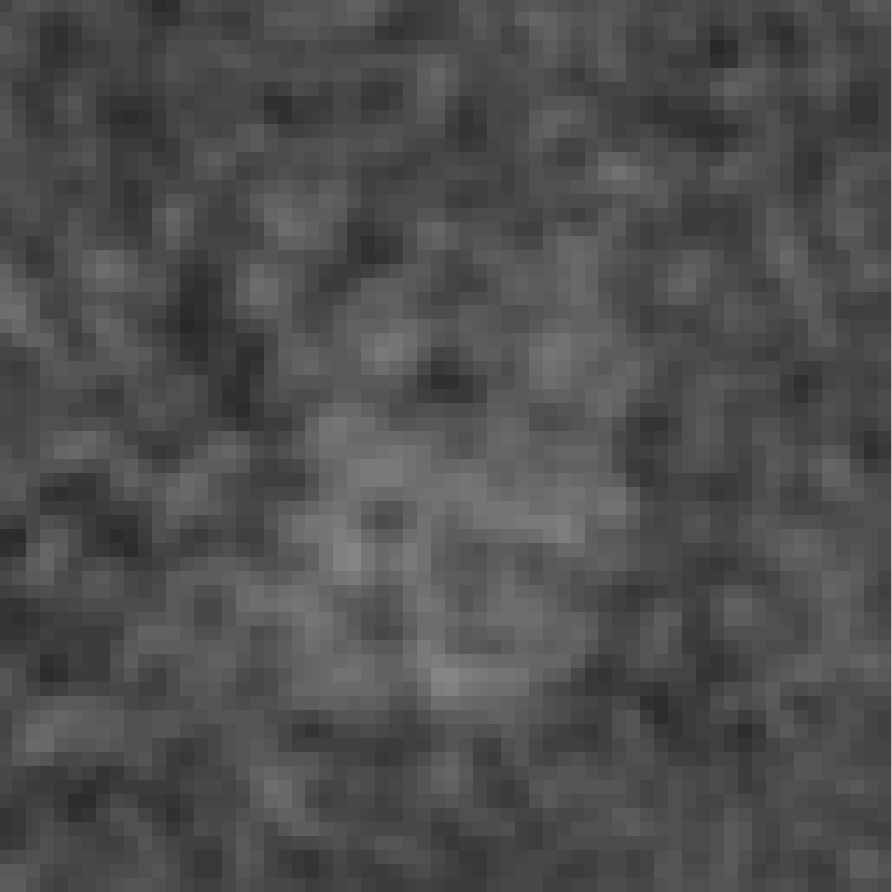}
}\\
 \subfigure[SNR=0.100, $\nu=0.3$]{\label{fig:noise_case4}
   \includegraphics[width=1.4in]{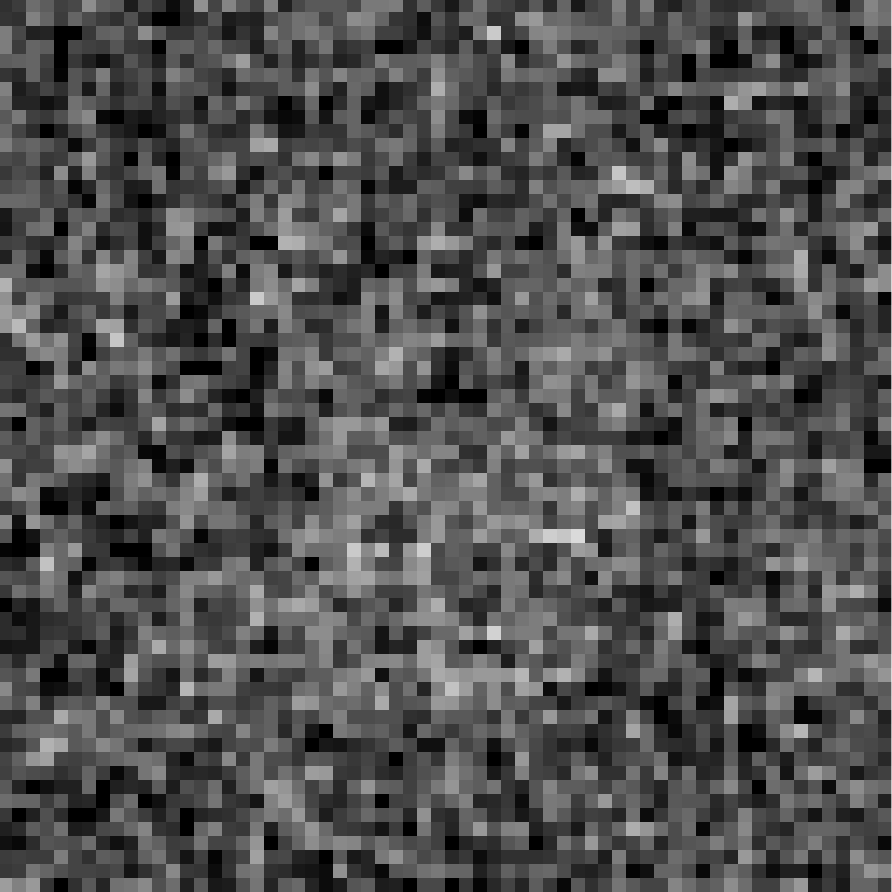}
}
 \subfigure[Blurred version of (g)]{\label{fig:blur_case4}
   \includegraphics[width=1.4in]{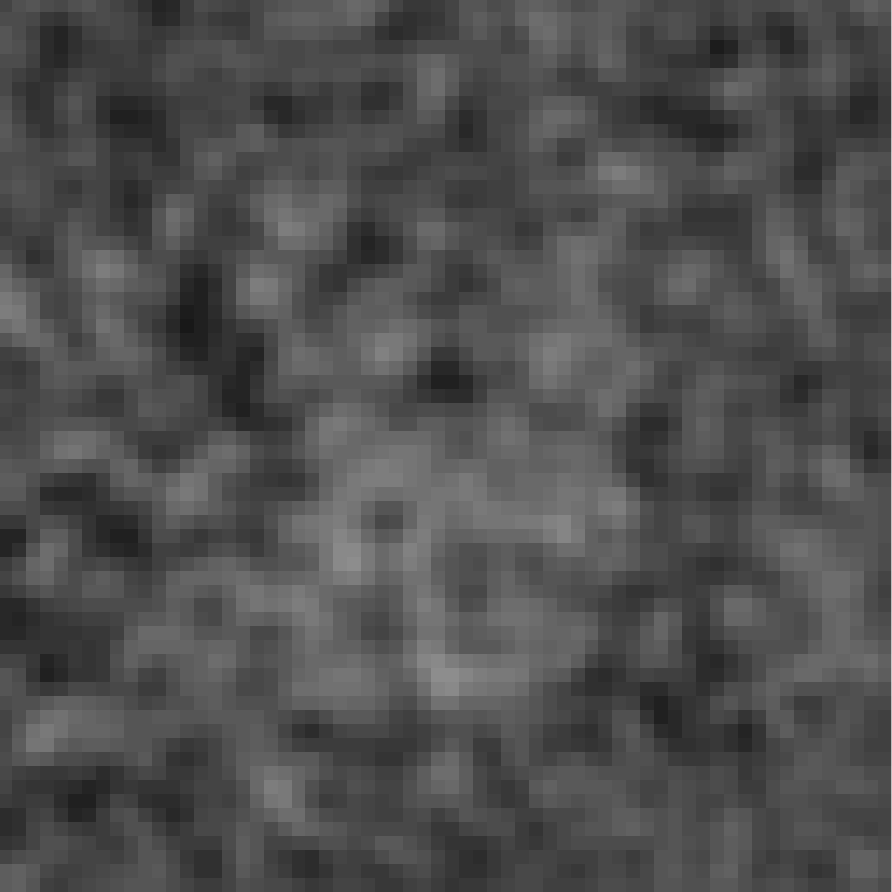}
}\\
  \caption{(a), (c), (e), and (g) Example test images in Case
  1,2,3, and 4, respectively. (b), (d), (f), and (h) The blurred
  version of (a), (c), (e), and (g), respectively. $\nu$ is the
  correlation between the noise in two adjacent pixels in the background.
    }\label{fig:noise_blur}
\end{figure}

\begin{table}
\centering
        \begin{tabular}{|c|c|c|c|c|}
  \hline
     & {\small Case 1} & {\small Case 2 }& {\small Case 3} & {\small Case 4 } \\
  \hline
   {\small SNR }& \hspace{4mm} 0.025  \hspace{4mm} & \hspace{4mm}
   0.050 \hspace{4mm} & \hspace{4mm} 0.100 \hspace{4mm} & \hspace{4mm}
   0.100 \hspace{4mm}  \\
     {\small Correlation }& \hspace{4mm} 0.0  \hspace{4mm} & \hspace{4mm}
   0.0 \hspace{4mm} & \hspace{4mm} 0.0  \hspace{4mm} & \hspace{4mm}
   0.3 \hspace{4mm}  \\
  \hline
\end{tabular}
 \caption{Signal-to-noise ratios and correlations of the adjacent
 noise pixels for four test cases.
 } \label{tab:cases}
\end{table}

\begin{figure}
  \centering
  \subfigure[CMPA alignment]{\label{fig:CMPA_0_case1}
  \includegraphics[width=1.6in]{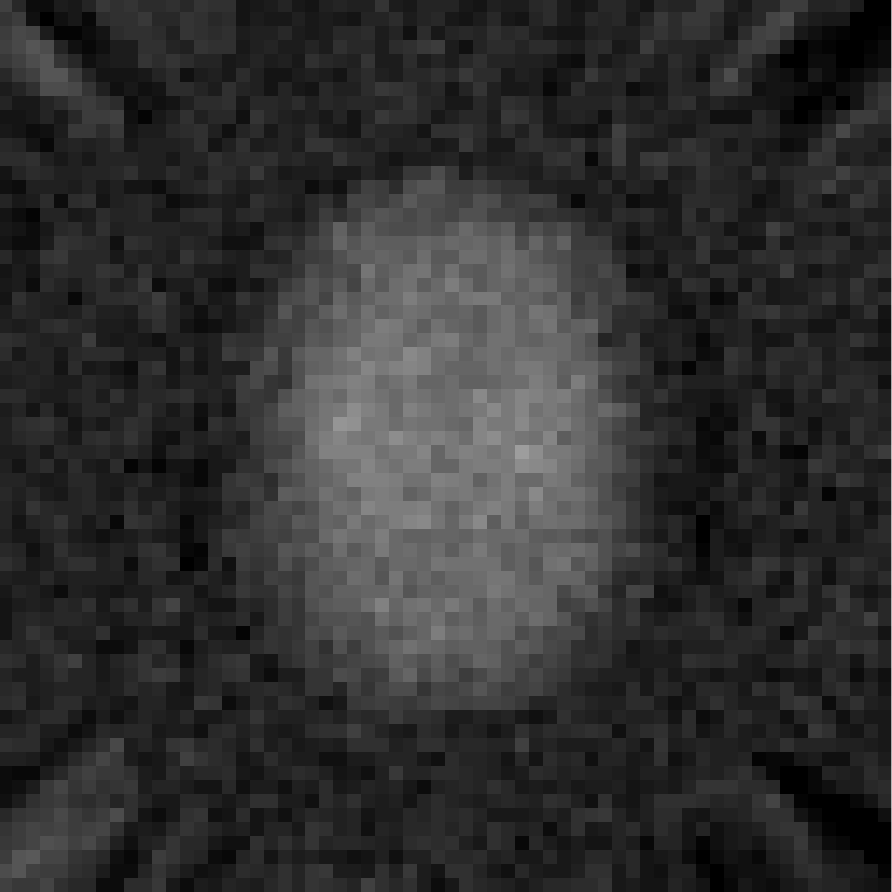}
  }
  \subfigure[\#1]{\label{fig:CMPA_1_case1}
  \includegraphics[width=1.6in]{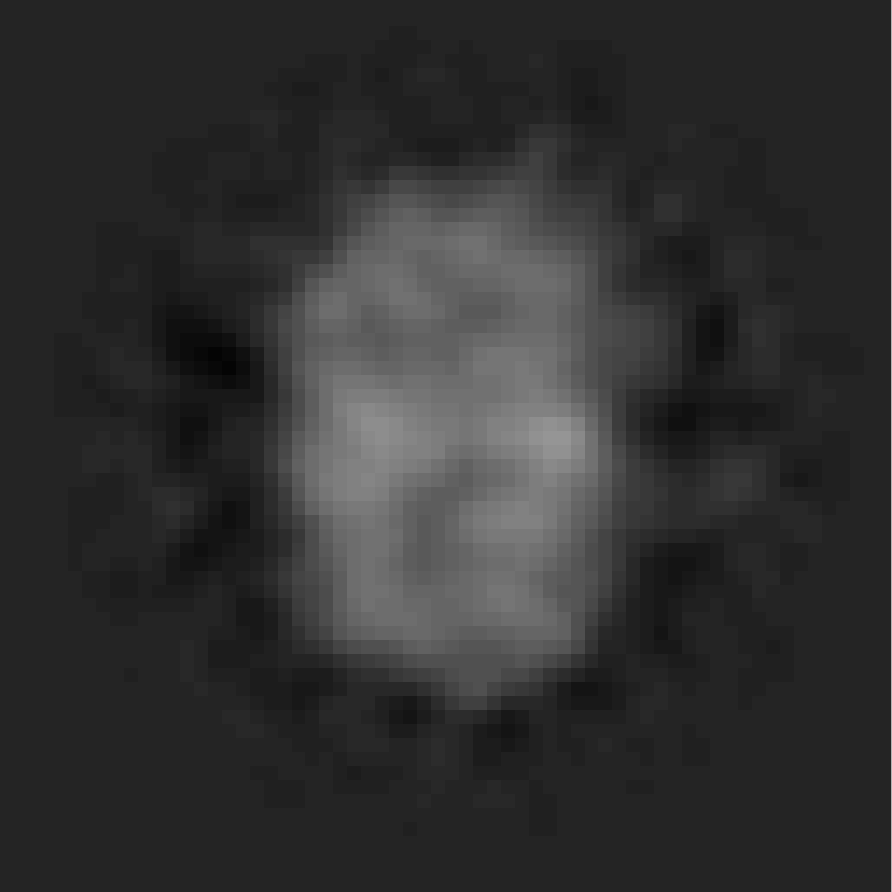}
  }
  \subfigure[\#10]{\label{fig:CMPA_2_case1}
  \includegraphics[width=1.6in]{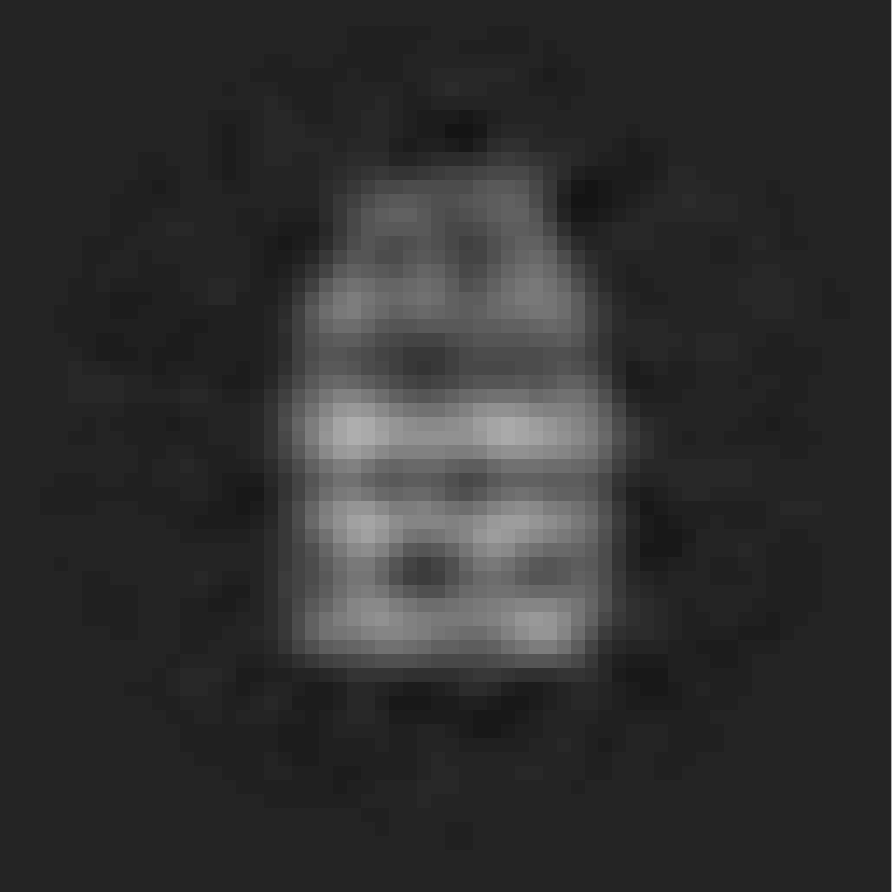}
  }\\
   \subfigure[\#11]{\label{fig:CMPA_3_case1}
  \includegraphics[width=1.6in]{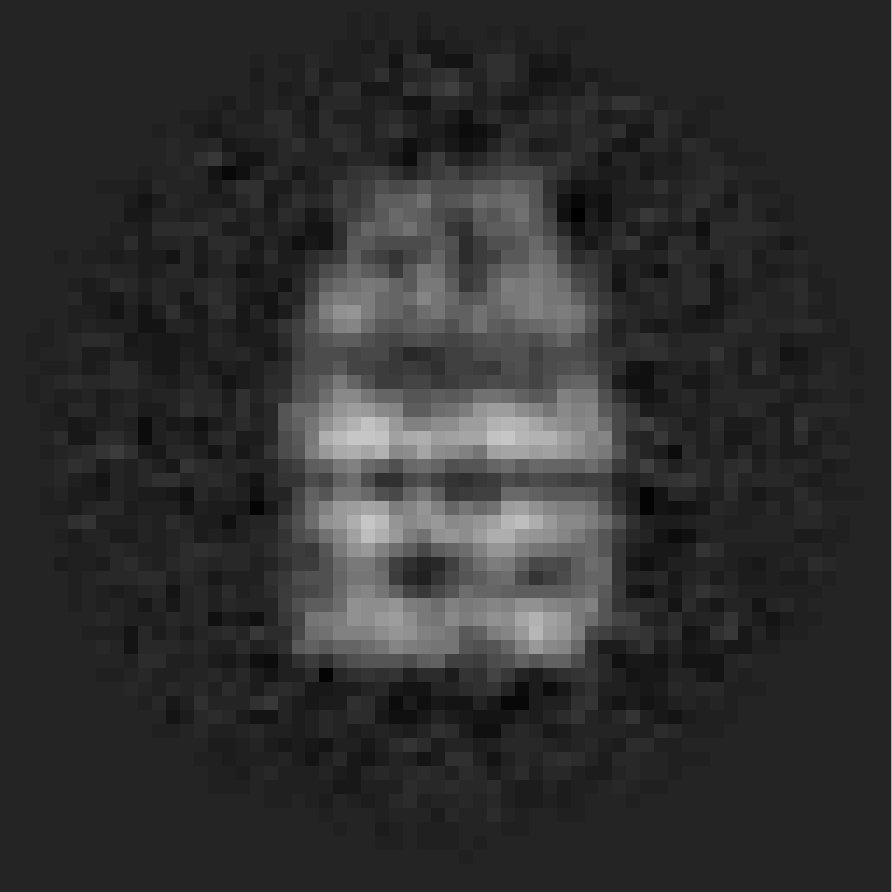}
  }
   \subfigure[\#19]{\label{fig:CMPA_4_case1}
  \includegraphics[width=1.6in]{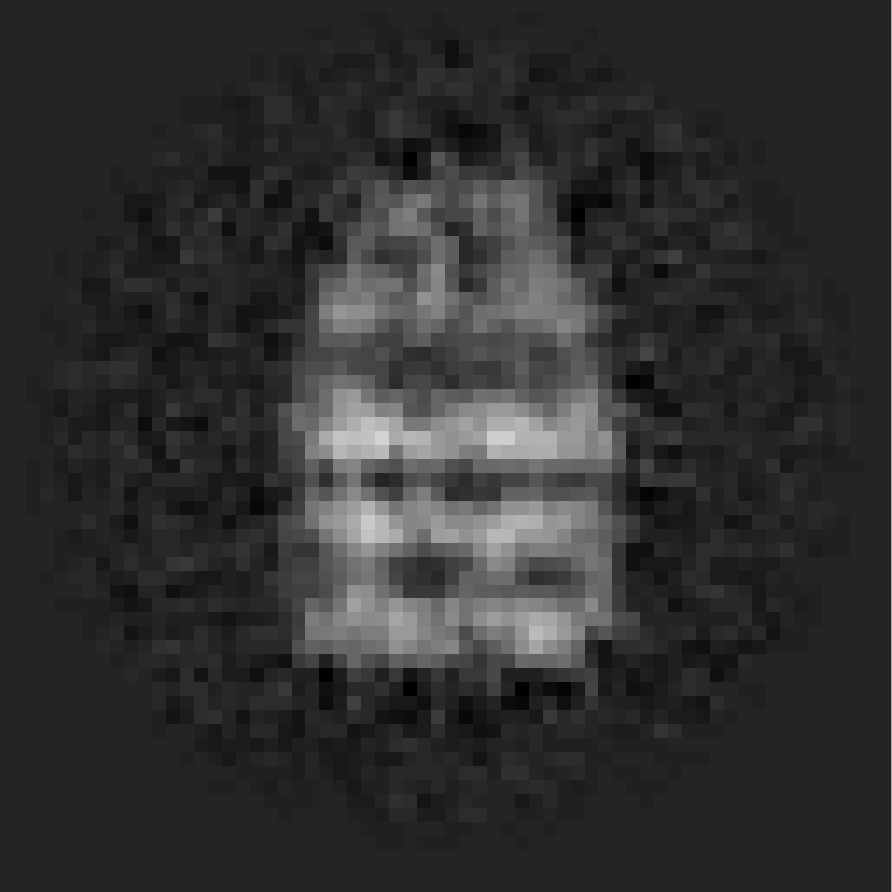}
  }
  \subfigure[\#30]{\label{fig:CMPA_5_case1}
  \includegraphics[width=1.6in]{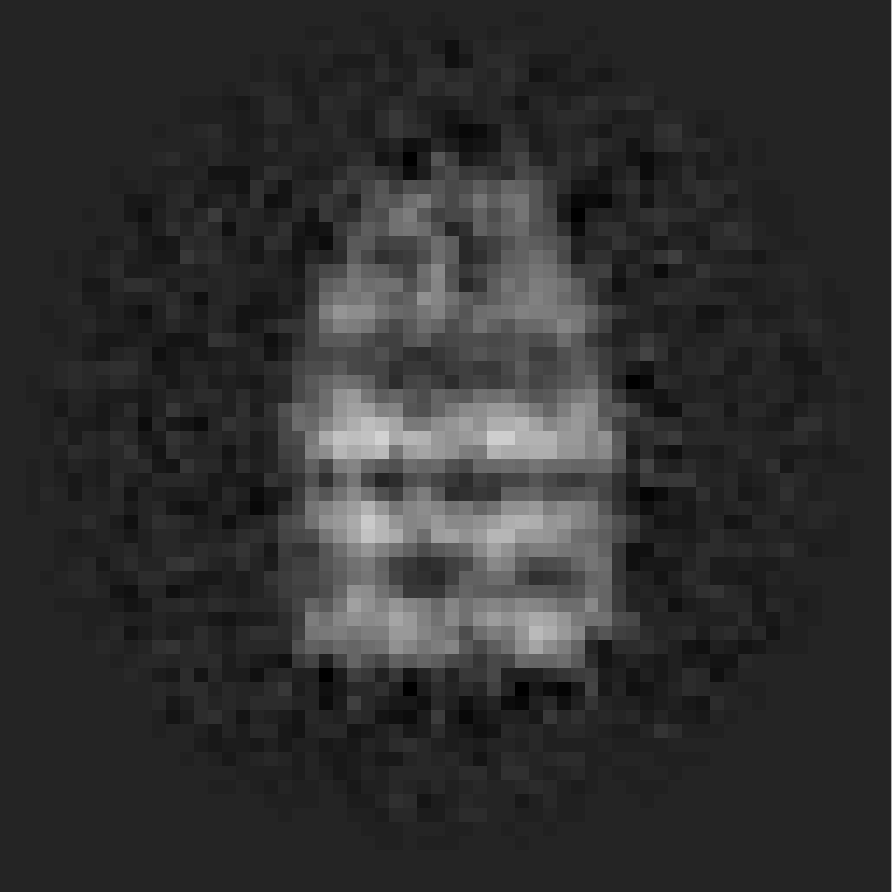}
  }
  \caption{The result of the new method for Case 1.
  (a) Result by CMPA (Phase 1) (b) Initial image for Phase 2
  (c) Result of Phase 2 (d) Initial image for Phase 3
  (e) Result of Phase 3 (f) Resulting image after 30 iterations. }\label{fig:Case1_our_method}
\end{figure}

\begin{figure}
  \centering
  \subfigure[Reference 1]{
  \includegraphics[width=1.6in]{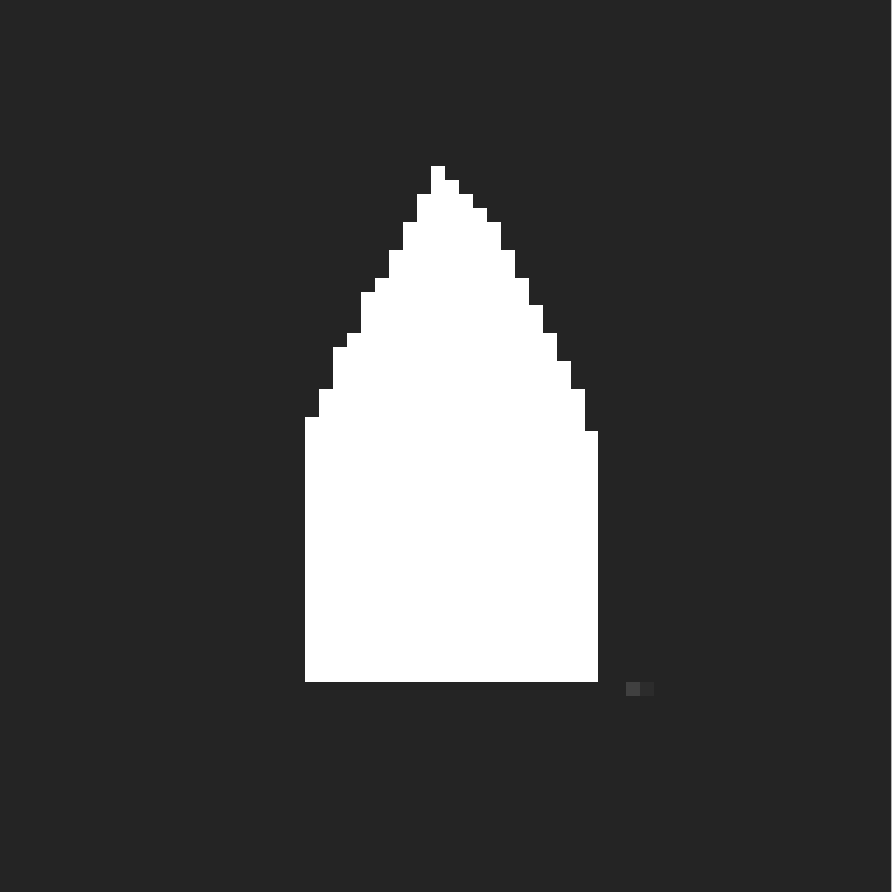}
  }
  \subfigure[Reference 2]{
  \includegraphics[width=1.6in]{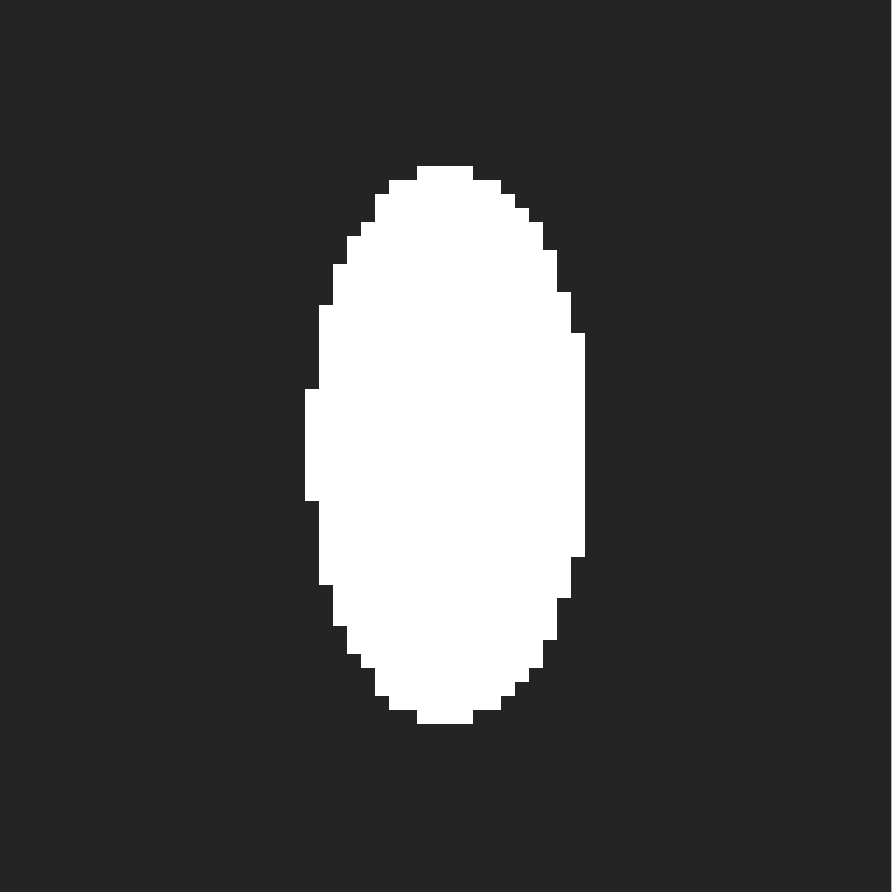}
  }
  \subfigure[Reference 3]{
  \includegraphics[width=1.6in]{noise_case1}
  }\\
   \subfigure[CC 1]{
  \includegraphics[width=1.6in]{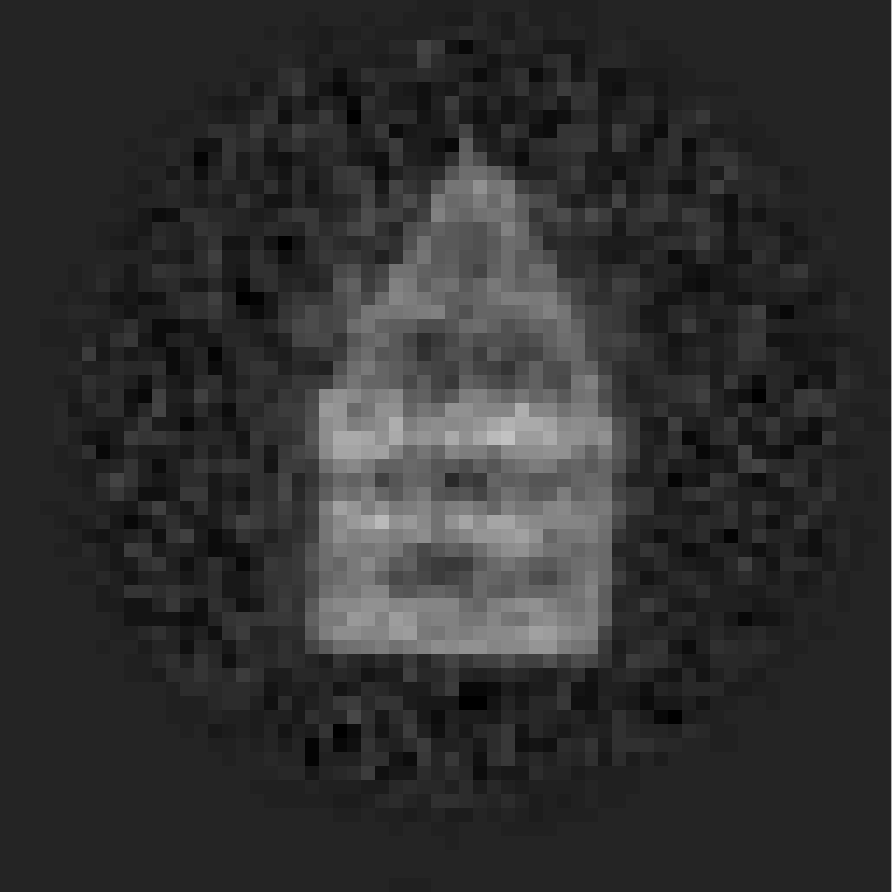}
  }
     \subfigure[CC 2]{
  \includegraphics[width=1.6in]{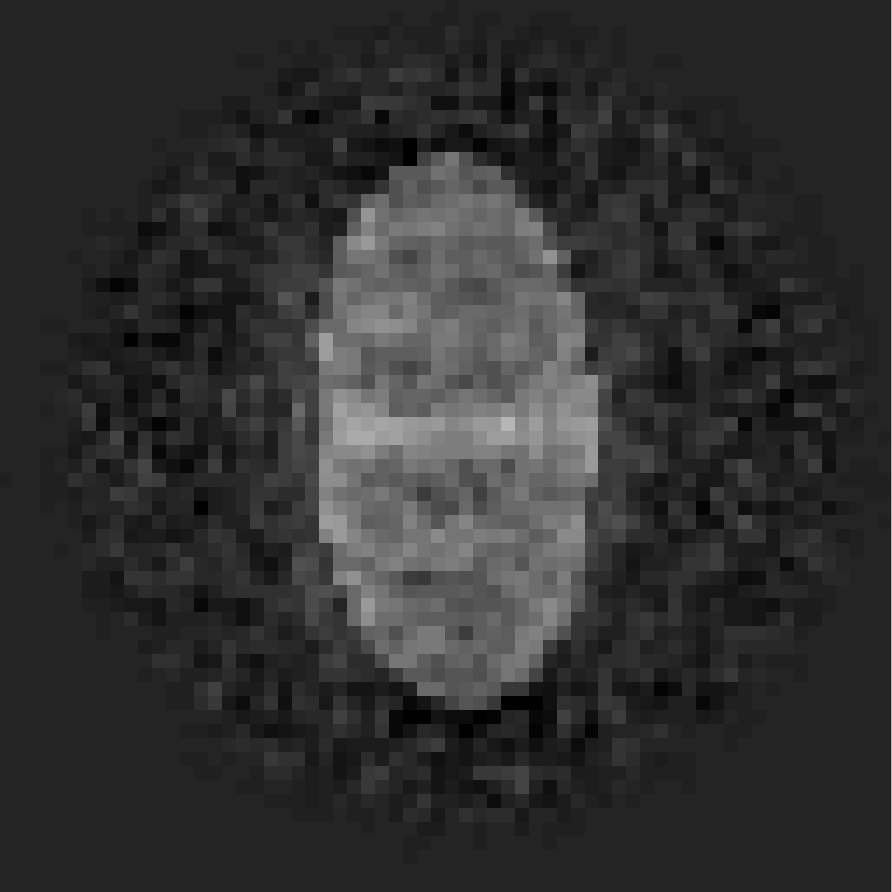}
  }
    \subfigure[CC 3]{
  \includegraphics[width=1.6in]{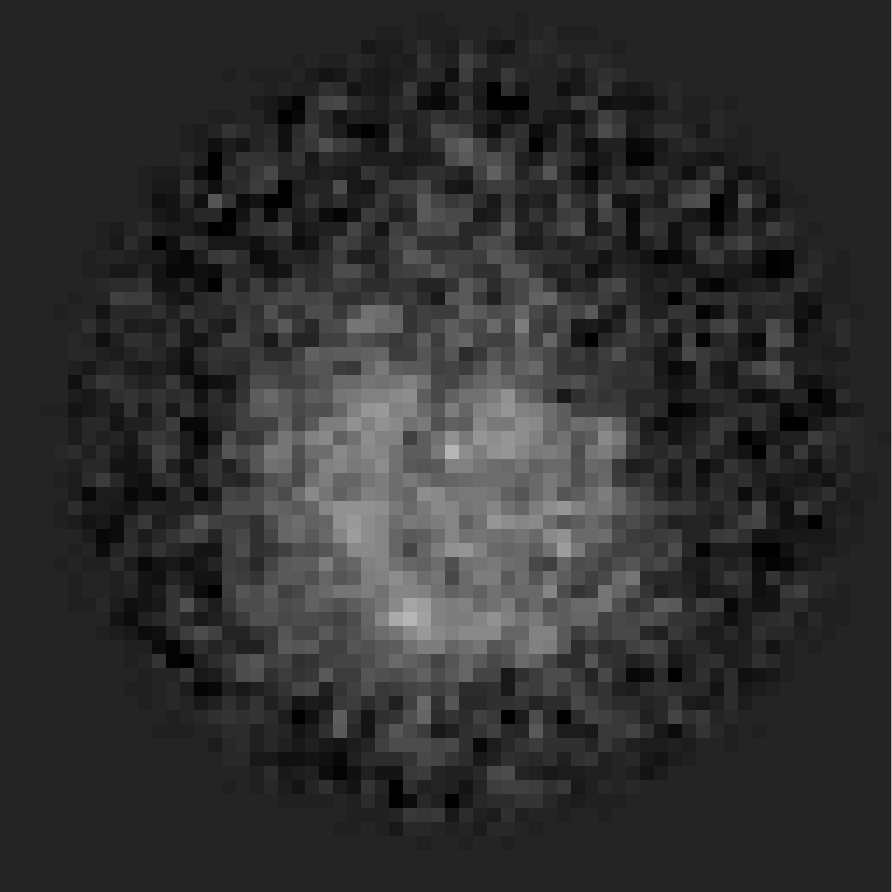}
  }\\
   \subfigure[ML 1]{
  \includegraphics[width=1.6in]{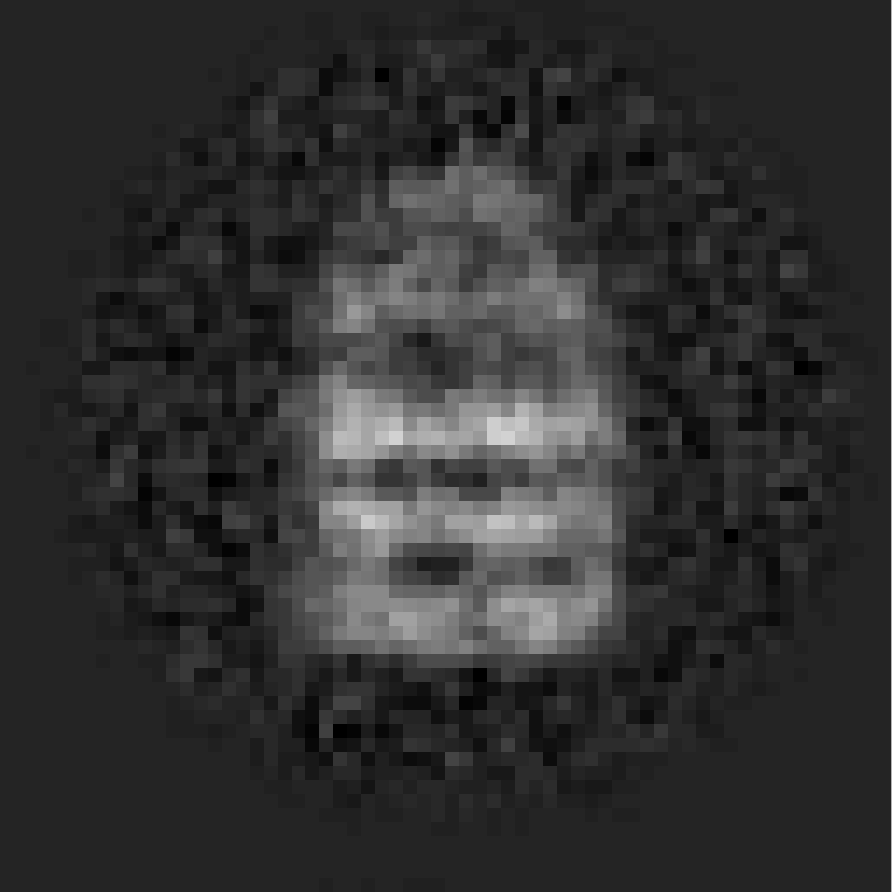}
  }
     \subfigure[ML 2]{
  \includegraphics[width=1.6in]{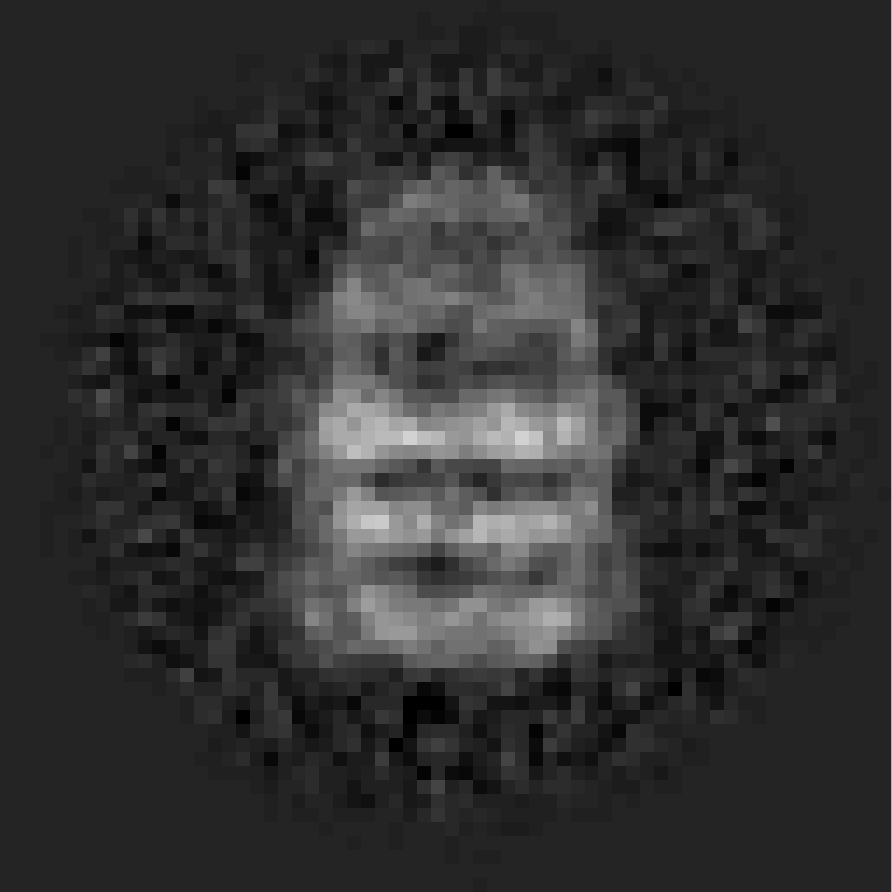}
  }
    \subfigure[ML 3]{
  \includegraphics[width=1.6in]{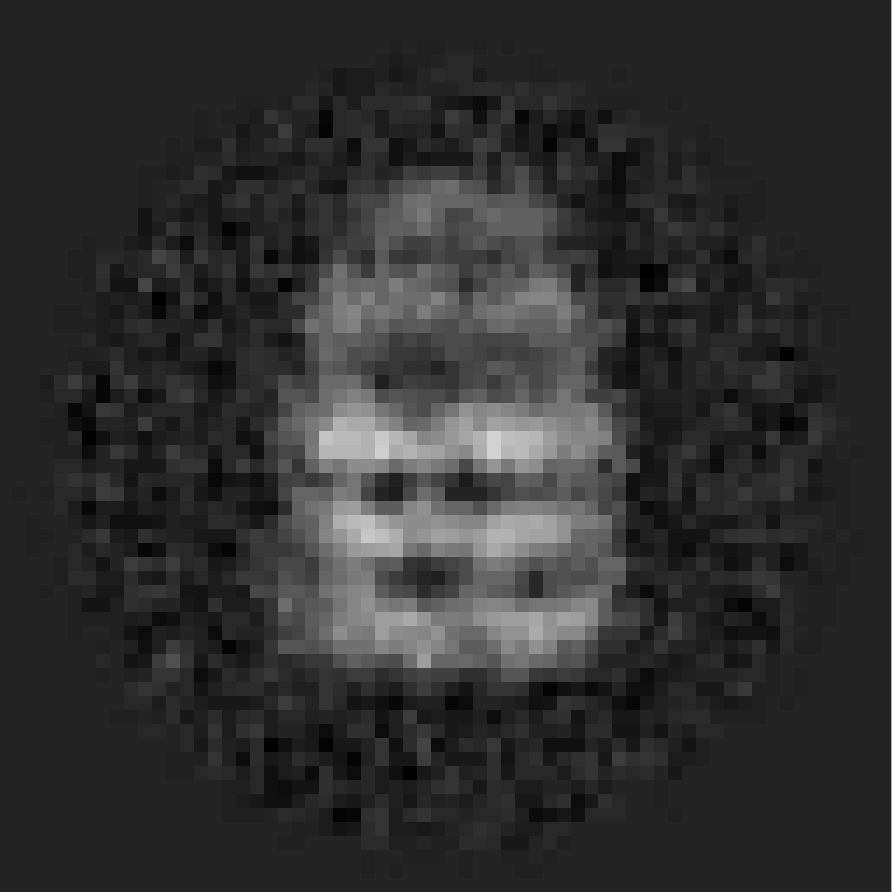}
  }
    \caption{The results of the conventional CC and ML methods for Case 1 with three reference images.
    Reference 3 is one class image.
    }\label{fig:case1_their_method}
\end{figure}

\begin{figure}
  \centering
   \subfigure[]{
   \includegraphics[width=2.5in ]{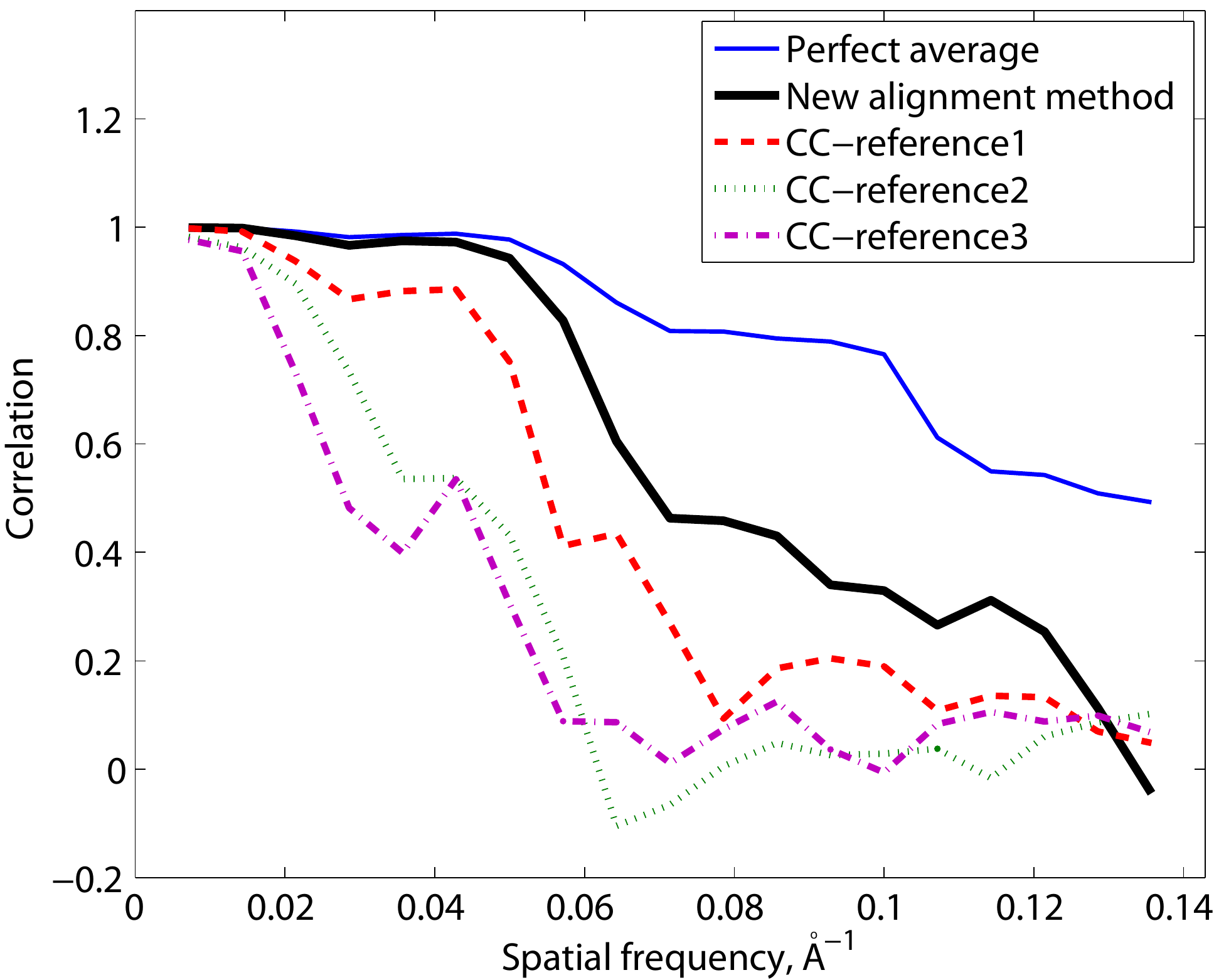}
}
 \subfigure[]{
  \includegraphics[width=2.5in]{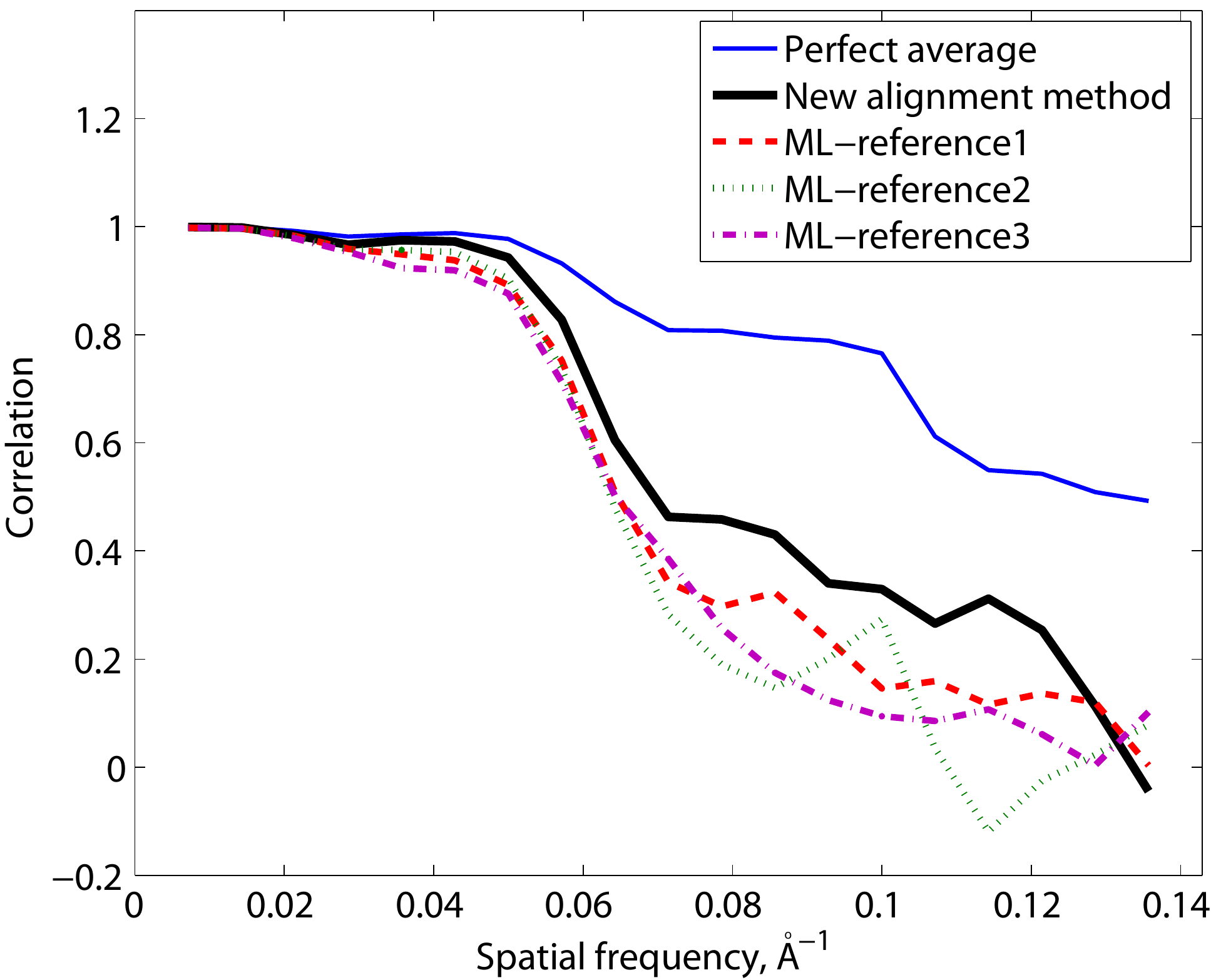}
  }
  \caption{FRC plots for Case 1 (a) Comparison of FRCs of the new
  method and the CC method (b) Comparison of FRCs of the new
  method and the ML method
  }\label{fig:FRC_case1}
\end{figure}

\begin{figure}
  \centering
   \includegraphics[width=3in]{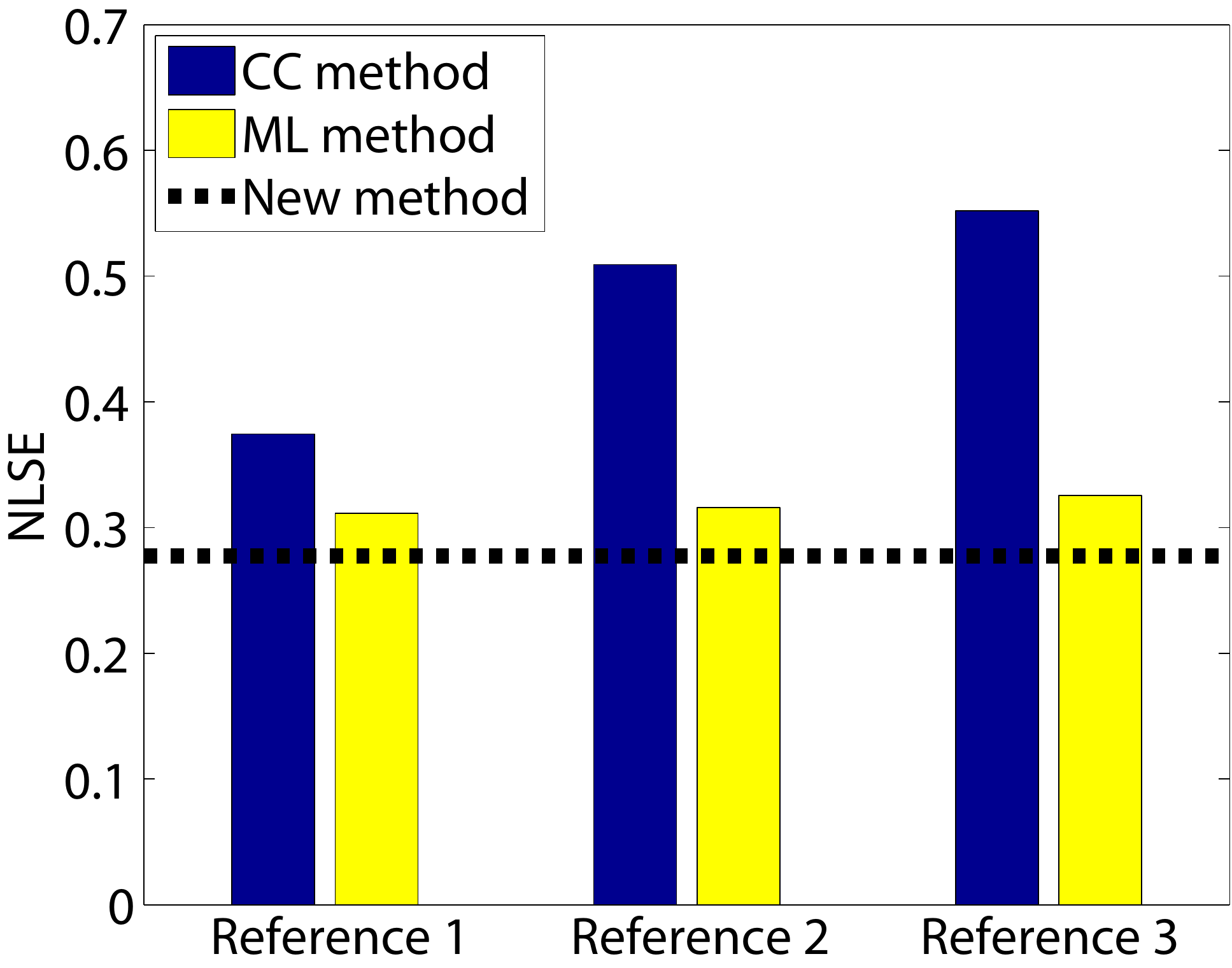}
  \caption{Image difference between the
   alignment/refinement results and the original
   image measured by NLSE for Case 1}\label{fig:difference_case1}
\end{figure}

\section{Results}
\label{sc:result}

In this section we compute the alignment and the class average for
 four cases defined in Table \ref{tab:cases} using the new method.

To generate the synthetic data images, we first transform (i.e.
translate and rotate) the clear projection image shown in Figure
\ref{fig:original}. The image size is $64\times 64$. The
rotational angles are sampled from a uniform distribution on $[0,
\,\, 2\pi)$. The translation distances are sampled from a Gaussian
distribution with the standard deviation, $5$ pixel. This setting
is consistent with the assumption in~\cite{sigworth1998maximum}.
After transforming, we add noise to the transformed projection.
The intensity of the noise is determined so that the resulting
image has the SNRs defined in Table~\ref{tab:cases}. The parameter
$\nu$ is the correlation coefficient between the noise in adjacent
pixels. The method of generating the noise with $\nu$ was
introduced in \cite{park:ijrr_stoch}. Figure
\ref{fig:perfect_average} shows the class average of 500 class
images with the perfect alignments for Case 1. Figure
\ref{fig:noise_blur} shows the noisy data images for the four
cases and their blurred version which is used in Phase 2 shown in
Figure \ref{fig:diagram}.

The search space for translation is bounded by $(-2.35
\xi_{\sigma}\,\,\,2.35 \xi_{\sigma})$. Since the probability
density function for the rotational angles is bimodal, the two
spaces $(-2.35 \xi_{\theta},\,\,\,2.35 \xi_{\theta})$ and $(-2.35
\xi_{\theta} +\pi,\,\,\,2.35\xi_{\theta}+\pi)$ are searched. Note
that $\xi_{\sigma}$ and $\xi_{\theta}$ were given in
(\ref{eq:sigma_define1}) and (\ref{eq:sigma_define2}). They are
computed from the background noise properties, and are not
adjustable parameters. The value 2.35 is the value dictated by
Gaussian statistics to guarantee that 98\% of the mass under the
Gaussian distribution is sampled. The translational misalignment
is limited to a multiple of one pixel length because translation
by sub-pixel distance involves interpolation and increases the
computation time without bringing new information out of images.
This limited search also enables us to compute the CC using the
DFT. We sample 22 angles for rotational search using the inverse
transform sampling. Two sets of 11 samples are drawn from the
intervals $(-2.35 \xi_{\theta},\,\,\,2.35 \xi_{\theta})$ and
$(-2.35 \xi_{\theta} +\pi,\,\,\,2.35\xi_{\theta}+\pi)$,
respectively.

Figure \ref{fig:CMPA_0_case1} shows the coarse alignment obtained
by the CMPA match for Case 1. In Phase 2 we use the blurred
version of class images to avoid false peaks in the cross
correlation. Even though the optimal parameter for the artificial
blurring is determined as $\sigma=0.5$ pixel if we apply the full
process of Phase 2 described in Section \ref{sc:flow}, we observe
that the final result after Phase 3 is not heavily dependent on
the blurring parameter as long as we consider $\sigma = 0.25,
0.50, 0.75$ or $1.00$ pixel. For demonstration we fix the standard
deviation for the artificial blurring as $\sigma = 1$ pixel
without losing the benefit of Phase 2. For Case 1, Figure
\ref{fig:CMPA_1_case1} shows the first iteration result in Phase
2. And the iteration in Phase 2 was repeated up to 10 iterations
(Figure \ref{fig:CMPA_2_case1}). From the $11^{th}$ iteration
(Figure \ref{fig:CMPA_3_case1}), Phase 3 is applied until it
converges. The $19^{th}$ iteration (Figure \ref{fig:CMPA_4_case1})
shows the converged result. As mentioned earlier, during
iterations, the combined transformations for each image are
computed and recorded.

Figure \ref{fig:case1_their_method} shows the results by the CC
method and the ML method for Case 1 with three different reference
images. For the fair comparison, we use 22 equally-spaced samples
on the interval $[0,\,\,\,2\pi)$ for angles in the CC method and
the ML method.

\begin{figure}
  \centering
  \subfigure{
  \includegraphics[width=1.6in]{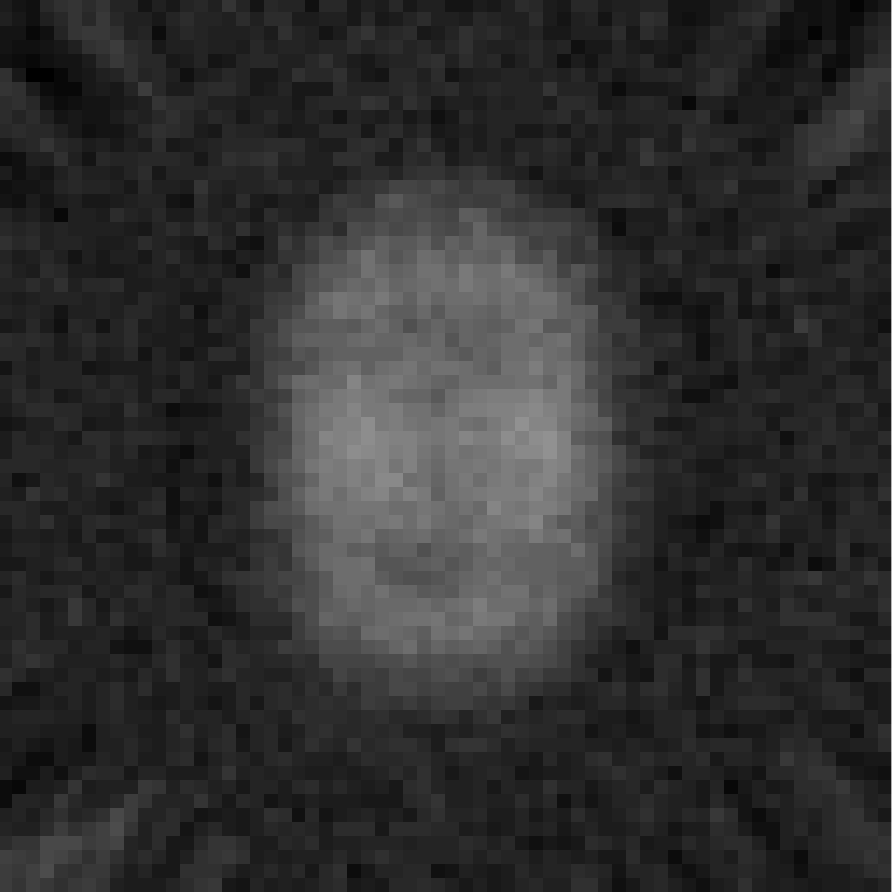}
  }
   \subfigure{
  \includegraphics[width=1.6in]{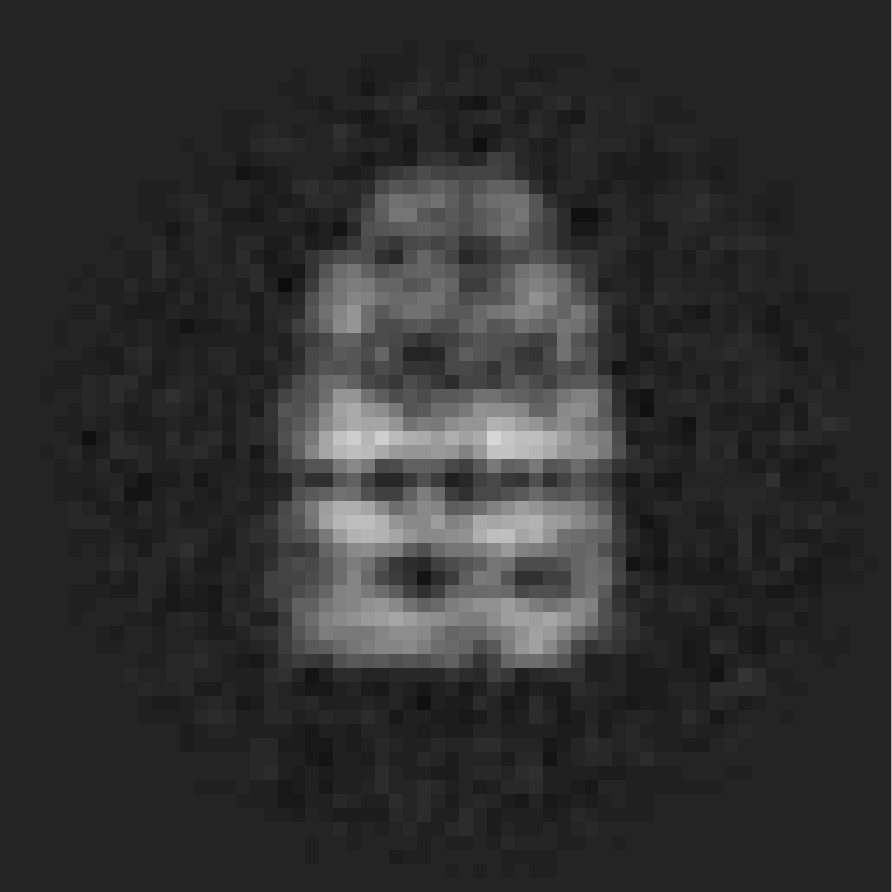}
  }
  \caption{The result of the new method for Case 2
  (a) CMPA alignment (b) Final result}\label{fig:Case2_our_method}
\end{figure}

\begin{figure}
  \centering
   \subfigure[]{
   \includegraphics[width=2.5in ]{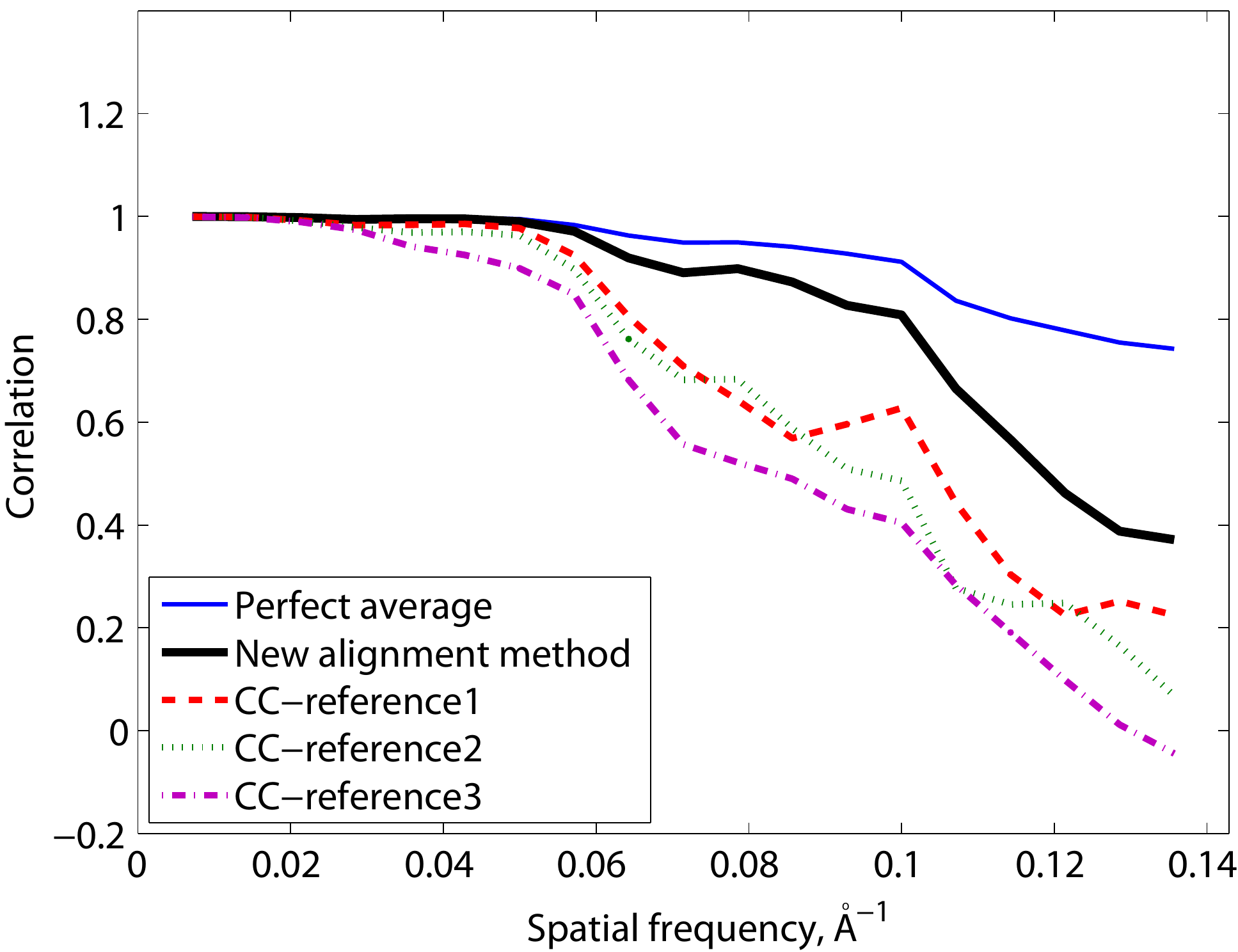}
}
 \subfigure[]{
  \includegraphics[width=2.5in]{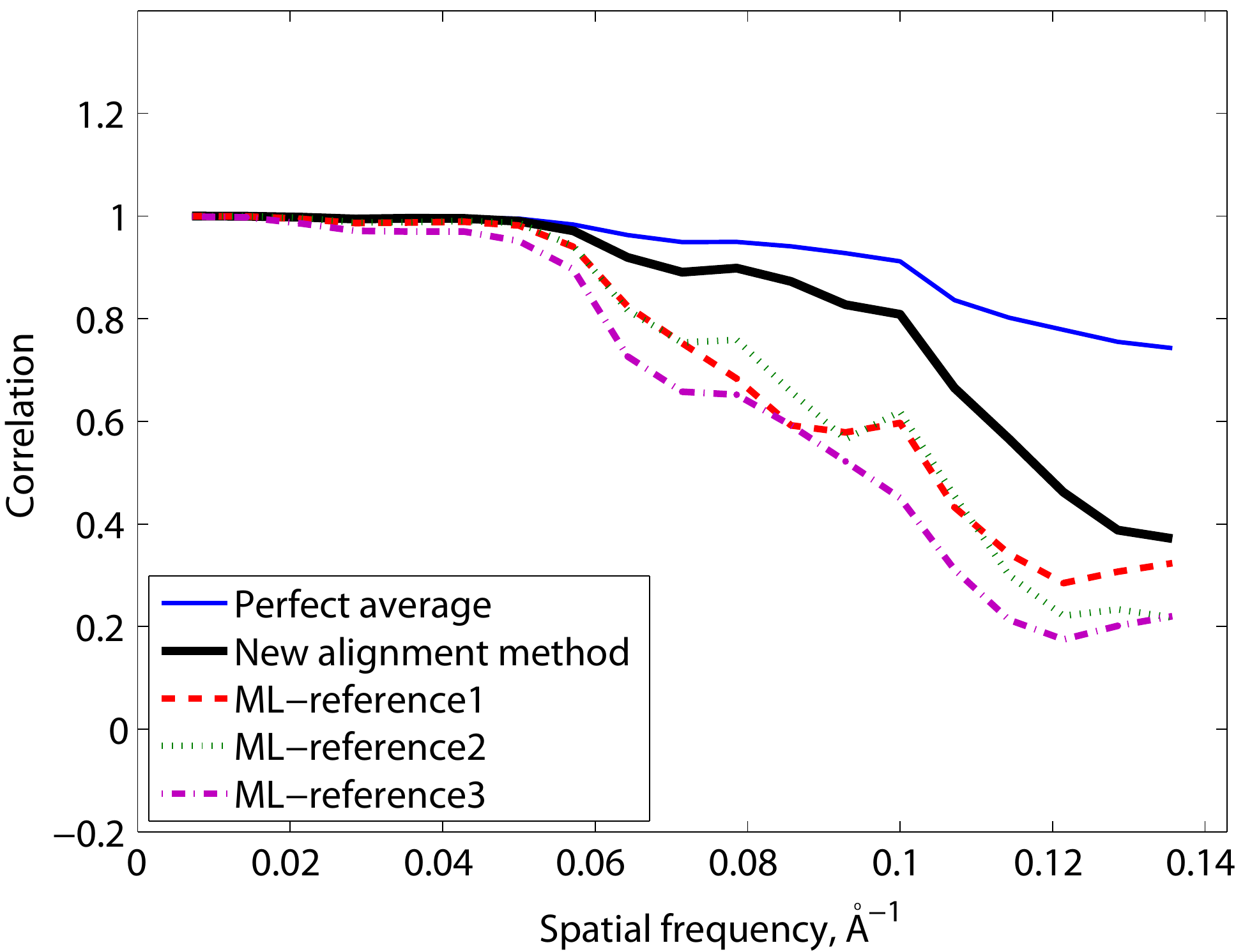}
  }
 \caption{FRC plots for Case 2 (a) Comparison of FRCs of the new
  method and the CC method (b) Comparison of FRCs of the new
  method and the ML method}\label{fig:FRC_case2}
\end{figure}

\begin{figure}
  \centering
   \includegraphics[width=3in]{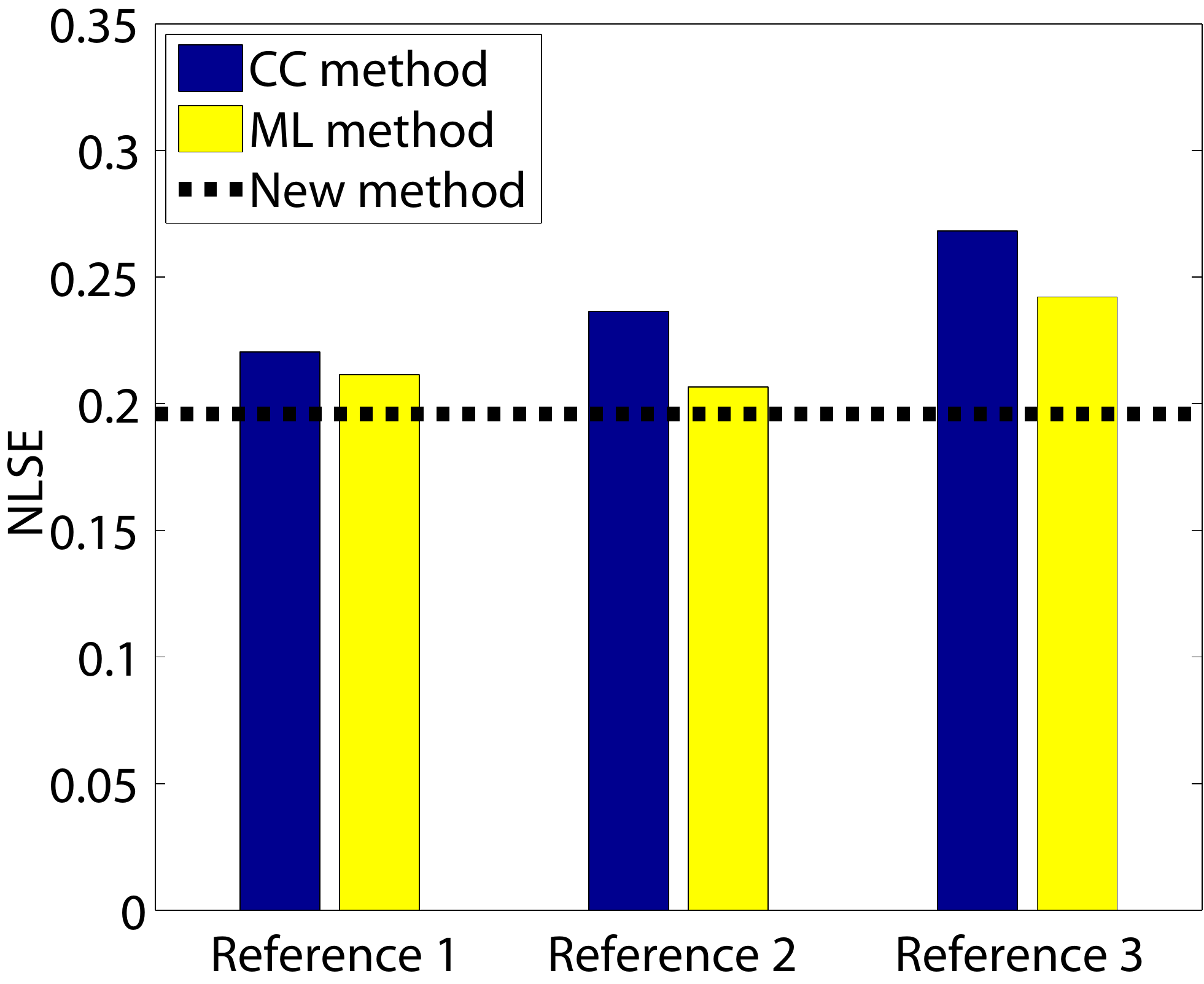}
  \caption{Image difference between the
   alignment/refinement results and the original
   image measured by NLSE for Case 2 }\label{fig:difference_case2}
\end{figure}

\begin{figure}
  \centering
   \subfigure[]{
   \includegraphics[width=2.5in ]{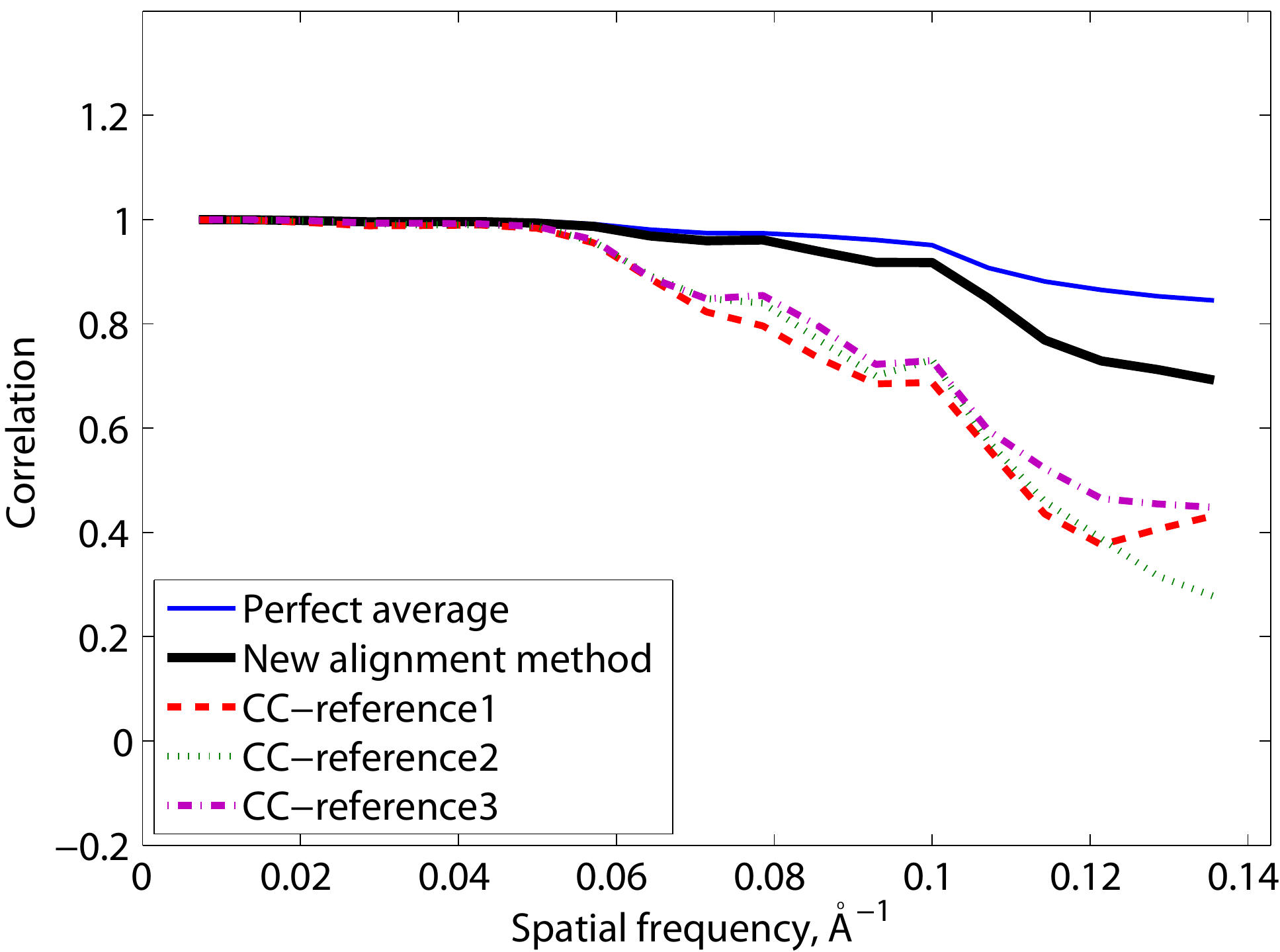}
}
 \subfigure[]{
  \includegraphics[width=2.5in]{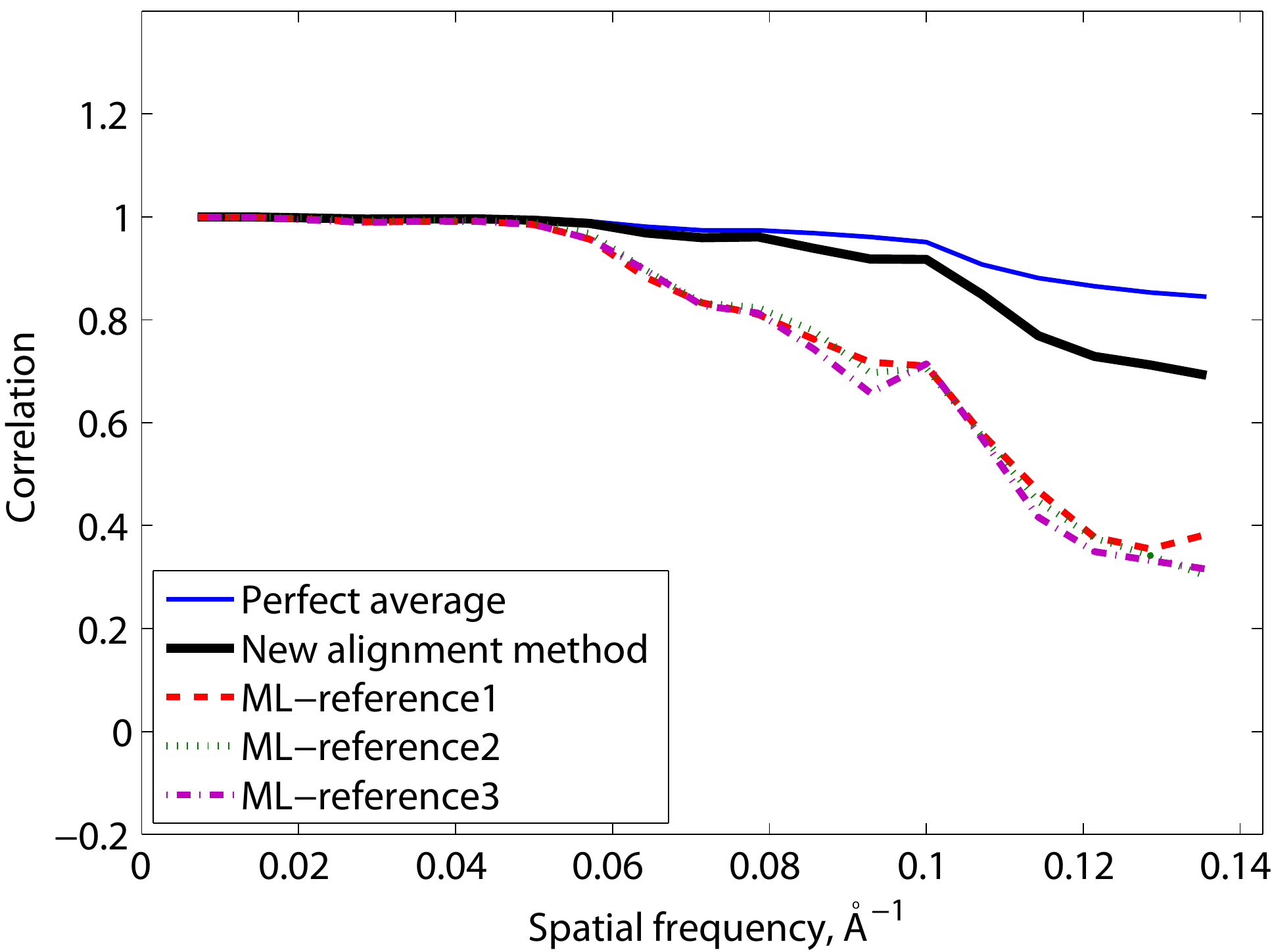}
  }
 \caption{FRC plots for Case 3 (a) Comparison of FRCs of the new
  method and the CC method (b) Comparison of FRCs of the new
  method and the ML method}\label{fig:FRC_case3}
\end{figure}

\begin{figure}
  \centering
   \includegraphics[width=3in]{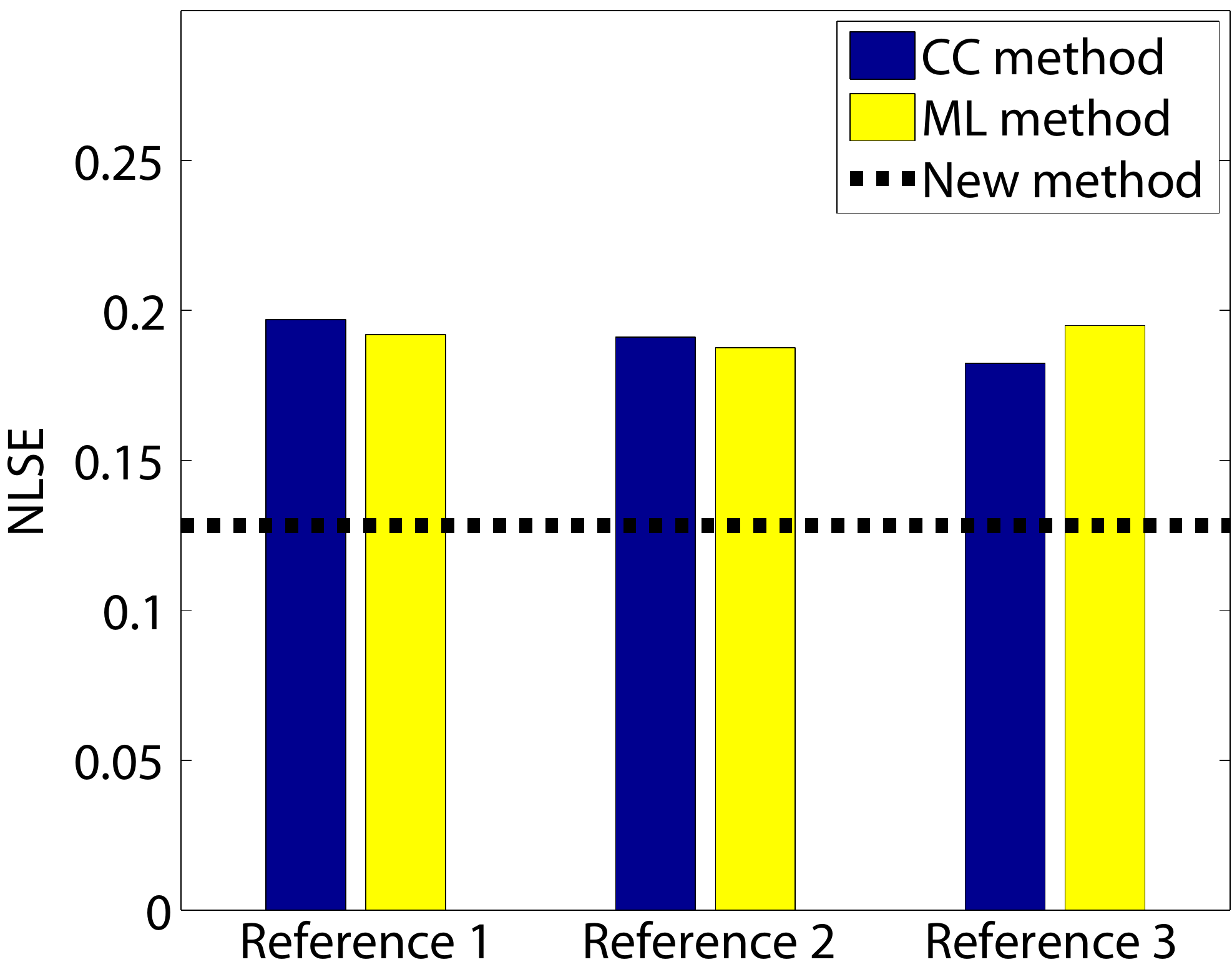}
 \caption{Image difference between the
   alignment/refinement results and the original
   image measured by NLSE for Case 3 }\label{fig:difference_case3}
\end{figure}

\begin{figure}
  \centering
   \subfigure[]{
   \includegraphics[width=2.5in ]{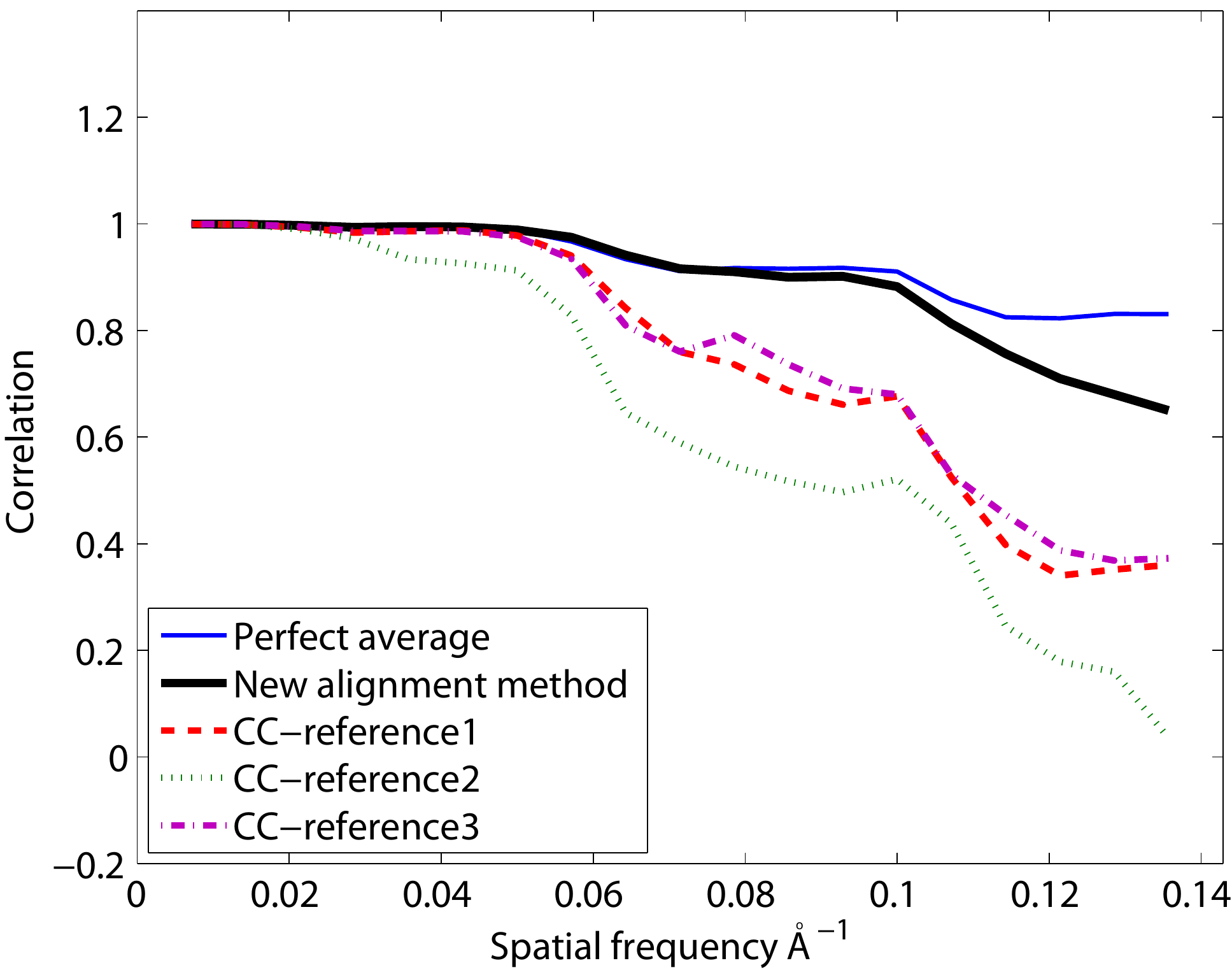}
}
 \subfigure[]{
  \includegraphics[width=2.5in]{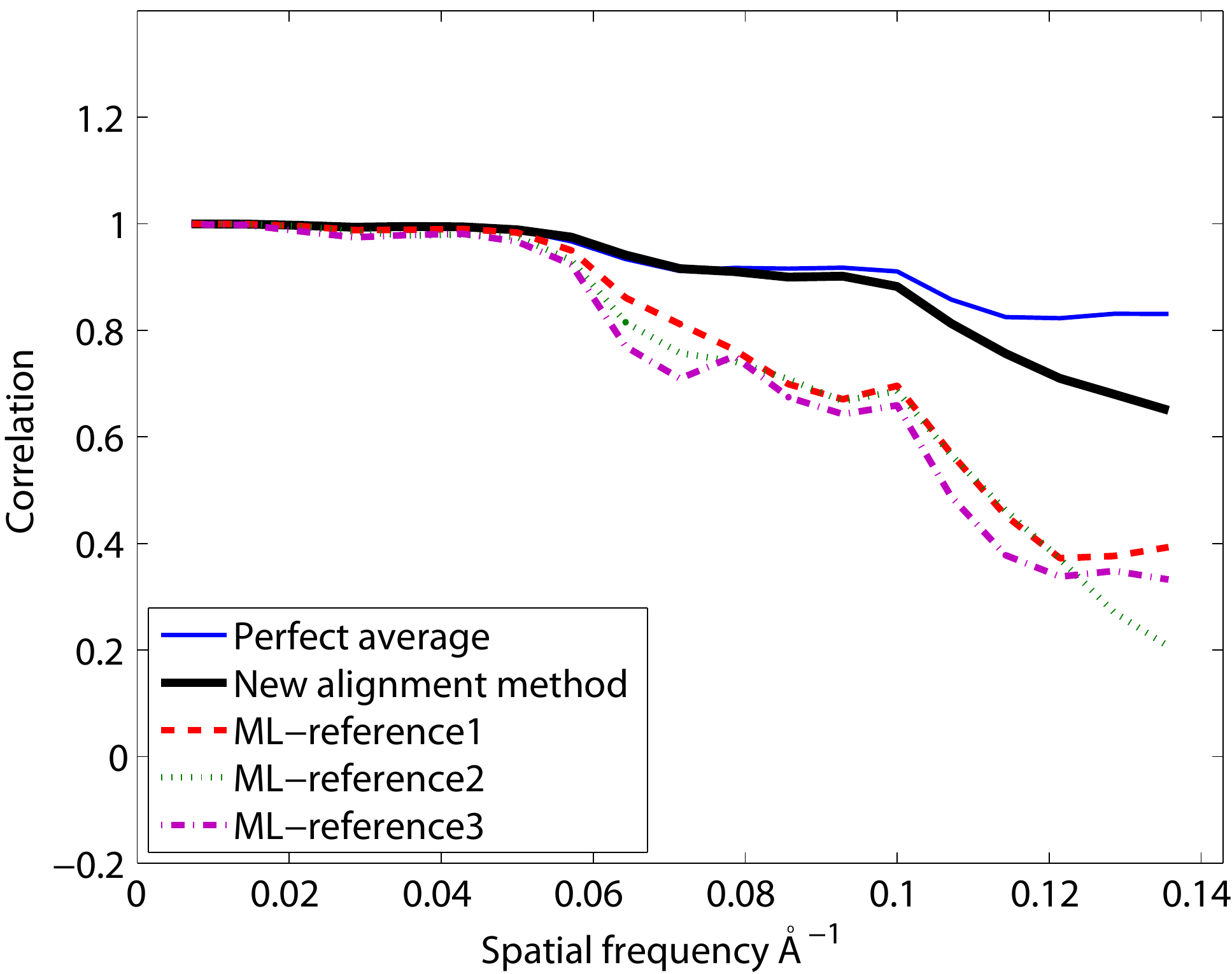}
  }
 \caption{FRC plots for Case 4 (a) Comparison of FRCs of the new
  method and the CC method (b) Comparison of FRCs of the new
  method and the ML method}\label{fig:FRC_case4}
\end{figure}

\begin{figure}
  \centering
   \includegraphics[width=3in]{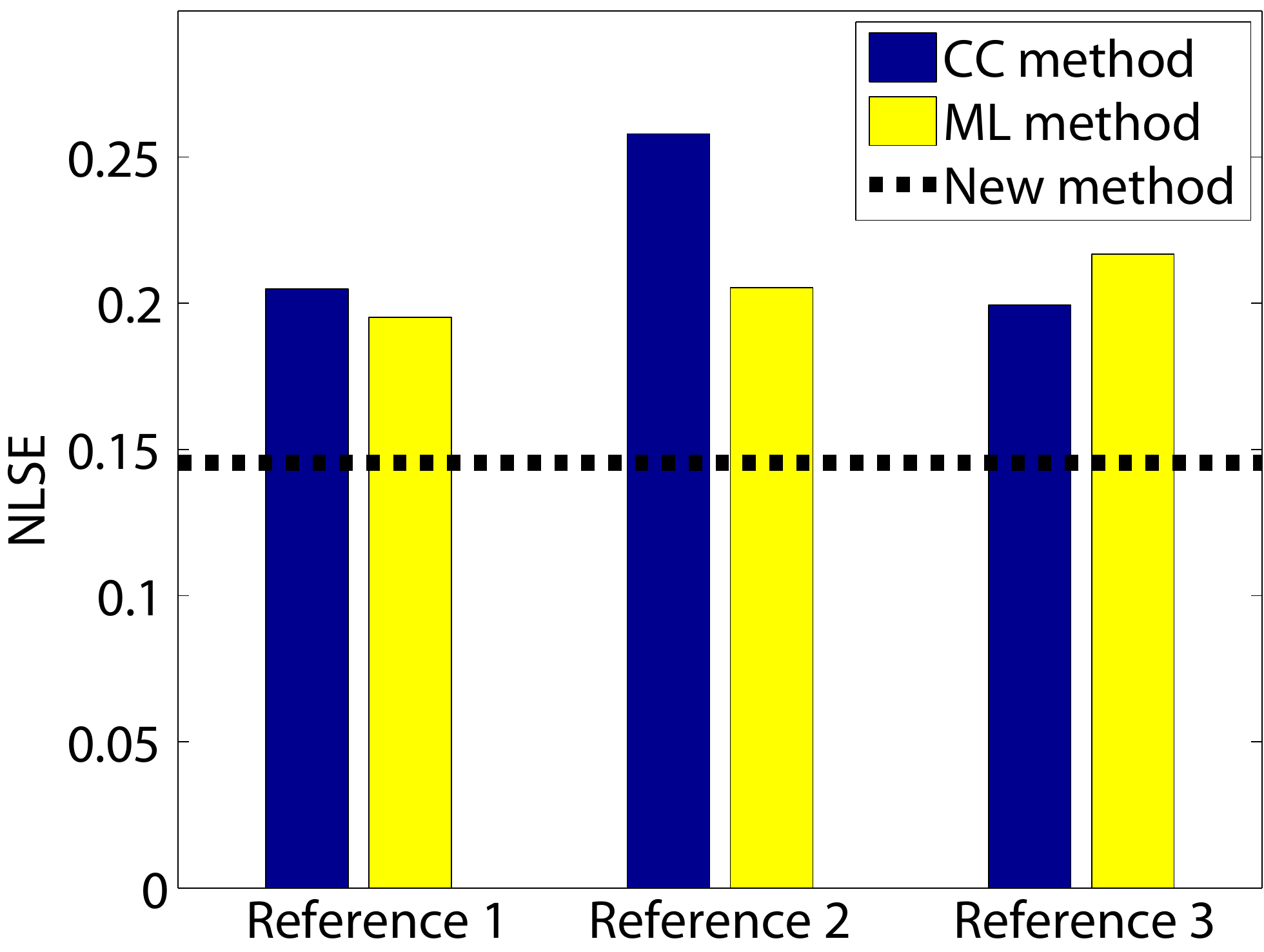}
  \caption{Image difference between the
   alignment/refinement results and the original
   image measured by NLSE for Case 4}\label{fig:difference_case4}
\end{figure}

Figure \ref{fig:FRC_case1} shows the Fourier ring correlation
(FRC) curves between the pristine projection shown in Figure
\ref{fig:original} and resulting images by our new method, the CC
method and the ML method. The FRC of the average with perfect
alignments shown in Figure \ref{fig:perfect_average} is also
shown. Figure \ref{fig:difference_case1} shows the image
differences between the projection shown in Figure
\ref{fig:original} and other resulting images by our new method,
the CC method and the ML method. The differences are measured
using the normalized lease-square error (NLSE). The NLSE of a image $U(m,n)$ relative
to another image $V(m,n)$, is defined as
$$
\mbox{NLSE}= \sqrt{\frac{\sum^N_{m=1}\sum^N_{n=1}
[U(m,n)-V(m,n)]^2 }{\sum^N_{m=1}\sum^N_{n=1} [V(m,n)]^2}}.
$$
As shown in Figure \ref{fig:FRC_case1} and
\ref{fig:difference_case1}, the resulting image obtained by our
new method is better than the other results. Figure
\ref{fig:Case2_our_method}-\ref{fig:difference_case2},
\ref{fig:FRC_case3}-\ref{fig:difference_case3}, and
\ref{fig:FRC_case4}-\ref{fig:difference_case4} show the resulting
images and their assessment for Case 2,3, and 4, respectively. The
resulting images for Case 3 and 4 are not presented since they
look similar to the other cases. The assessments for the results
are more important and they are provided in Figure
\ref{fig:FRC_case3}-\ref{fig:difference_case4}. They consistently
show that our new method performs better than existing ones.

When we compute the FRC and the normalized least squared errors, we align two images before
computation, because similarity and difference between two images
are sensitive to their alignment. Since two images that we compare
here are a underlying clear image and a resulting class average by
alignment methods, we can apply the cross correlation method to
align them without concern about false peaks in cross correlation
of noisy images. For more accurate alignment for image comparison,
we also apply the image segmentation method to eliminate the area
of the residual noise in the class average.

While in Figure \ref{fig:FRC_case2}
the FRC curves of the results of the new
method are better than those of the existing methods over all the
frequency range, Figure \ref{fig:FRC_case1} shows that the curve
of the result of the new method is lower than the other curves at
the highest frequency. This does not mean that the resulting image
of the new method is worse than the others, because the curve of
the result of the new method is higher at the other frequency and
the image difference shown in Figure \ref{fig:difference_case1}
supports the fact that the new method produces a better image.

In these tests, we used 500 images for one class. The
pre-alignment for the 500 images by matching CMPA took
approximately 20 seconds using a conventional PC. One iteration in
Phase 2 and 3 shown in Figure~\ref{fig:diagram} took about 4.4
seconds. The number of iterations until convergence for each test
case is shown in Table \ref{tab:num_iter}. One iteration in the
classical CC and ML methods takes approximately 2.6 and 6.0
seconds, respectively. The number of iterations until convergence
in these existing method are also shown in Table
\ref{tab:num_iter}. The total computation time of the new method
for each case in Table \ref{tab:cases} is about 100 seconds, which
is the similar computation time of the ML method. The conventional
CC method takes about 50 seconds until convergence, but the
resulting images are not good as measured using FRC and normalized
least squared error. It is
important to note that the preprocess to compute or generate a
starting reference image for the conventional CC and ML methods is
not considered in this computation time. Therefore the total
computation time will be increased if that process is included.

\begin{table}
\centering
        \begin{tabular}{|c|c|c|c|c|c|c|c|}
  \hline
     & {\small New method} & {\small CC 1 }& {\small CC 2} & {\small CC 3}
     & {\small ML 1 }& {\small ML 2} & {\small ML 3} \\
  \hline
   {\small Case 1}& 19 &17&14&20&25&26&25\\
    {\small Case 2}& 23&11&12&11&27&27&20\\
     {\small Case 3}& 13&12&19&13&24&21&15\\
      {\small Case 4}& 17&11&16&10&26&22&28\\
  \hline
\end{tabular}
 \caption{Number of iterations for convergence} \label{tab:num_iter}
\end{table}

%

%

%

\section{Conclusion}\label{sc:conclusion}
In this work, we developed a new alignment method for class
averaging in single particle electron microscopy. The new method
consists of two steps: pre-alignment and re-alignment. In the
pre-alignment process, images in a class are aligned using their
centers of mass and principal axes. Although this pre-alignment
does not generate an accurate alignment, it provides a reasonable
staring point for the next re-alignment process. Furthermore, we
can quantitatively characterize the distribution of misalignments
in this pre-alignment method. In the second step, we re-align the
images using the results from the first step. Essentially we apply
the CC method to re-align images from the first step with the
reduced search space that was created based on the statistics of
misalignment. In order to avoid problems related to false peaks in
the cross correlation, blurred version of the images are used in
the first phase of the second step. After iteration with the
blurred images, we use the original image to find more accurate
alignment.

The pre-alignment step using the CMPA method has several technical
benefits. First, it produces a data-driven reference image for the
following iteration. Second, the resulting alignment and the
corresponding averages are independent of the order of input
images unlike the previous work in \cite{penczek1992three}. Third,
it provides the distribution of misalignment regardless of the
initial distribution of poses of the projection. In the ML method,
the initial distribution of the position of the projection is
assumed to be Gaussian, and the orientations are assumed to be
uniformly random. What if the initial orientational distributions
are different from uniform and how can the distribution be
estimated to ensure that this assumption is correct? Our
pre-alignment step replaces the pre-existing distribution of the
pose of the projection with one that is known. Though the
resulting alignment is not perfect, the statistics of misalignment
can be quantified and accurately estimated. This misalignment is
also independent of the initial distribution of the pose of the
projection. Finally, a reduced search space based on the
distribution of misalignment can be obtained, and this leads to
more accurate alignment.

As we modified the conventional CC method using the CMPA matching
and the reduced search space, we can combine the ML method and the
CMPA matching method. We expect that the initial estimation for
the parameters in (\ref{eq:f_exp}) is possible. Therefore the
iteration in the ML method will converge faster. Technically the
distribution function (\ref{eq:f_exp}) should be modified to
reflect that the rotational misalignment after the CMPA matching
is no longer uniform. Specifically the function (\ref{eq:f_exp})
will be a form of (\ref{eq:pdf_misalignment}) and better
discretization of integral in (\ref{eq:L_func}) based on the new
distribution will result in better performance. We leave this work
for future research.

\section*{Acknowledgements}
This work was supported by NIH Grant R01GM075310. The authors
thank Dr. Fred Sigworth for providing the source code that was
developed in \cite{sigworth1998maximum}.





\bibliographystyle{elsarticle-num}
\bibliography{bibs}







\end{document}